\documentclass[twocolumn,epj]{svjour}  
\usepackage{graphicx}
\usepackage{multirow}
\usepackage{amsmath,amssymb,amsfonts}
\usepackage{mathrsfs}
\usepackage[title]{appendix}
\usepackage{xcolor}
\usepackage{textcomp}
\usepackage{manyfoot}
\usepackage{booktabs}
\usepackage{algorithm}
\usepackage{algorithmicx}
\usepackage{algpseudocode}
\usepackage{listings}
\usepackage[colorlinks,citecolor=blue,urlcolor=blue,linkcolor=blue]{hyperref}
\usepackage[numbers,sort&compress]{natbib}
\usepackage{accents}

\newlength{\dhatheight}
\newcommand{\doublehat}[1]{%
    \settoheight{\dhatheight}{\ensuremath{\hat{#1}}}%
    \addtolength{\dhatheight}{-0.25ex}%
    \hat{\vphantom{\rule{1pt}{\dhatheight}}%
    \smash{\hat{#1}}}}
\journalname{Eur. Phys. J. C}

\begin{document}
\pagenumbering{arabic}

\title{Higher-order asymptotic corrections and their application to the Gamma Variance Model}

\author{Enzo Canonero\inst{1} \href{https://orcid.org/0000-0002-7180-4562}{\includegraphics[scale=0.06]{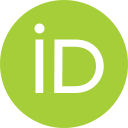}}
        \and
        Alessandra Rosalba Brazzale\inst{2} \href{https://orcid.org/0000-0003-1989-2697}{\includegraphics[scale=0.06]{orcid.png}}
        \and
        Glen Cowan\inst{1} \href{https://orcid.org/0000-0001-8363-9827}{\includegraphics[scale=0.06]{orcid.png}}
}
\institute{Physics Department, Royal Holloway, University of London, U.K. 
           \and
           Department of Statistical Sciences, University of Padova, Italy 
}

\date{Received: date / Accepted: date}

\abstract{
We present improved methods for calculating confidence intervals and $p$-values in situations where standard asymptotic approaches fail due to small sample sizes. We apply these techniques to a specific class of statistical model that can incorporate uncertainties in parameters that themselves represent uncertainties (informally, ``errors on errors'') called the Gamma Variance Model.  This model contains fixed parameters, generically denoted by $\varepsilon$, that represent the relative uncertainties in estimates of standard deviations of Gaussian distributed measurements.  If the $\varepsilon$ parameters are small, one can construct confidence intervals and $p$-values using standard asymptotic methods.  This is formally similar to the familiar situation of a large data sample, in which estimators for all adjustable parameters have Gaussian distributions. Here we address the important case where the $\varepsilon$ parameters are not small and as a consequence the first-order asymptotic distributions do not represent a good approximation. We investigate improved test statistics based on the technology of higher-order asymptotics (modified likelihood root and Bartlett correction).  The effective application of higher-order corrections removes an important computational barrier to the use of the Gamma Variance Model.
}


\maketitle


\section{Introduction}
\label{sec:intro}

In experimental sciences such as Particle Physics one collects data, 
here denoted by $\boldsymbol{y}$, and seeks to make inferences about a
hypothesis $H$ that defines the probability distribution for the data,
$P(\boldsymbol{y}|H)$. Often $P(\boldsymbol{y}|H)$ is indexed by a set
of \textit{parameters of interest} $\boldsymbol{\mu}$ and by a set of
\textit{nuisance parameters} $\boldsymbol{\theta}$, thus
$P(\boldsymbol{y}|H)=P(\boldsymbol{y}|\boldsymbol{\mu},
\boldsymbol{\theta})$.  The parameters of interest are the main objective of the analysis, whereas nuisance parameters are often
introduced to account for systematic uncertainties in the model.

We focus here on frequentist tests of the hypothesized parameters that
use a test statistic derived from the likelihood function $L(\boldsymbol{\mu},
\boldsymbol{\theta})=P(\boldsymbol{y}|\boldsymbol{\mu},
\boldsymbol{\theta})$.  These tests lead to confidence intervals or
regions for the parameters of interest as well as $p$-values that
quantify goodness of fit.  To find these results, one
requires the sampling distribution of test statistics that are
obtained from the likelihood function and are described in greater
detail below.  For appropriately defined test statistics, the
corresponding distributions can often be found using asymptotic
results based on theorems due to Wilks \cite{bib:Wilks1938} and Wald
\cite{bib:Wald1943} (see, e.g., \cite{bib:Cowan2011, bib:Algeri2019}).  The asymptotic
distributions are valid in specific limits, which usually correspond
to having a large data sample, whose size we will denote generically by $n$. 

In this paper we are interested specifically in the case where $n$ 
is not sufficiently large for the asymptotic distributions
of the relevant test statistics to represent a good approximation.  
In such problems one could use Monte Carlo methods to obtain the 
distributions, but this involves additional time-consuming computation.
Instead, one can modify the test statistic using the higher-order
asymptotic methods, specifically, the $r^*$ statistic of 
Barndorff-Nielsen \cite{bib:Barndorff-Nielsen1980} and the Bartlett 
correction \cite{bib:Bartlett1937}, as described, e.g., in
\cite{bib:Brazzale2007,bib:Cordeiro2014}.  With these methods, the
distribution of the modified statistic becomes closer to the asymptotic 
form, allowing one to find confidence intervals and $p$-values without
use of Monte Carlo.

In this paper we consider applications of higher-order asymptotic
methods to the Gamma Variance Model (GVM), which was proposed in
Ref.~\cite{bib:Cowan2019}.  In the GVM, measured values are modeled as
following Gaussian distributions with a mean that depends on the
parameters of the problem, and with variances $\sigma^2$ whose values
are themselves not certain.  The variances as well are thus taken as
adjustable parameters, and the values one would assign to them are
treated as measurements that follow a gamma distribution with
parameters $\alpha$ and $\beta$ (see Sec.~\ref{sec:gammaModel_def} 
below).  These parameters are assigned so that 
the gamma distribution's relative width
reflects the desired uncertainty on $\sigma^2$.  This is quantified
using the quantity $\varepsilon = 1/2\sqrt{\alpha}$, which to first
approximation is the relative uncertainty on the estimate of the
standard deviation $\sigma$, informally referred to as the ``error on
the error''.\footnote{In earlier references, e.g., \cite{bib:Cowan2019,bib:Cowan2022}, the
parameter $\varepsilon$ was denoted by $r$.}

In the Gamma Variance Model there is a correspondence between the
error-on-error parameters $\varepsilon$ and an effective sample size $n$ of
\begin{equation}
\label{eq:rdefinition}
n = 1 + \frac{1}{2 \varepsilon^2} \;.
\end{equation}

That is, the large-sample limit corresponds to the case
where $\varepsilon \rightarrow 0$ and thus the values of $\sigma$ are accurately
estimated.  For many analyses, however, the assigned values of
standard deviations for individual measurements may easily be
uncertain at the level of several tens of percent or more.  
In this case the effective sample size is
low and thus the asymptotic distributions of likelihood-based test
statistics are not necessarily valid.  The goal of this paper is to
apply higher-order asymptotics to this model and thus achieve more
accurate confidence levels and $p$-values.

In Sec.~\ref{sec:first_order_theory} we briefly review the basic 
techniques for finding confidence intervals and $p$-values in a general
likelihood-based analysis and  Sec.~\ref{sec:higher_order_asymp} 
describes how these techniques can be improved using higher-order asymptotics.   In Sec.~\ref{sec:gammaModel_def} 
we recall the important properties of the Gamma Variance Model, and then we explore use of higher-order asymptotic corrections to three specific realizations of the model:
in Sec.~\ref{sec:smm} we apply corrections to a simple example of the GVM based on a single measurement, in Sec.~\ref{sec:sam} to an average of measurements and in Sec.~\ref{sec:GVM_ave} to an average of measured values that includes control measurements to constrain nuisance parameters. A summary and conclusions are given in Sec.~\ref{sec:conclusions}.

\section{Parameter inference using the profile likelihood ratio}
\label{sec:first_order_theory}

In this section we review the basic technology used to find confidence intervals and $p$-values from test statistics derived from the likelihood ratio and the likelihood root by using the first-order asymptotic distributions based on Wilks' theorem.  Further details on these methods as applied in Particle Physics analyses can be found, e.g., in Ref.\cite{bib:Cowan2011}.

In statistical data analysis, the central object needed to carry out inference related to the parameters of interest $\boldsymbol{\mu}$ using measured data $\boldsymbol{y}$ is the likelihood function: $L(\boldsymbol{\mu}, \boldsymbol{\theta}) = P(\boldsymbol{y}|\boldsymbol{\mu}, \boldsymbol{\theta})$.  Suppose
there are $M$ parameters of interest $\boldsymbol{\mu} = (\mu_1, \ldots, \mu_M)$ and $N$ nuisance parameters $\boldsymbol{\theta} = (\theta_1, \ldots, \theta_N)$, which are introduced to account for systematic uncertainties. In frequentist statistics, a test of hypothesized parameter values can be carried out by defining a test statistic based on the (profile) likelihood ratio
\begin{equation}
\label{eq:lik_ratio}
    w_{\boldsymbol{\mu}} = -2\log\frac{L(\boldsymbol{\mu}, \doublehat{\boldsymbol{\theta}})}{L(\hat{\boldsymbol{\mu}}, \hat{\boldsymbol{\theta}})} = 2\left[\ell(\hat{\mu}, \hat{\boldsymbol{\theta}}) - \ell(\mu, \doublehat{\boldsymbol{\theta}})\right]\,.
\end{equation}
Here $\hat{\boldsymbol{\mu}}$ and $\hat{\boldsymbol{\theta}}$ are the Maximum Likelihood Estimators (MLEs) for the parameters of interest and the nuisance parameters, respectively, and $\doublehat{\boldsymbol{\theta}}$ are the profiled (or constrained) estimators of the nuisance parameters, given by the values of $\boldsymbol{\theta}$ that maximize the likelihood for a fixed value of $\boldsymbol{\mu}$ (here and below we use $\ell$ to denote the log-likelihood). 
The likelihood ratio is used to test the compatibility of a value of $\boldsymbol{\mu}$ with the experimental data, with greater $w_{\boldsymbol{\mu}}$ corresponding to increasing incompatibility.

The likelihood ratio can be used to derive a confidence region for the parameters of interest $\boldsymbol{\mu}$ (or a confidence interval if there is just one parameter of interest) by computing the $p$-value for a hypothesized value of $\boldsymbol{\mu}$,
\begin{equation}
\label{eq:pvalue}
    p_{\boldsymbol{\mu}} = \int_{w_{\boldsymbol{\mu}, \text{obs}}}^{\infty}f(w_{\boldsymbol{\mu}}|\boldsymbol{\mu}, \boldsymbol{\theta})dw_{\boldsymbol{\mu}}= 1 - F[w_{\boldsymbol{\mu}, \text{obs}}]\,.
\end{equation}
Here $f(w_{\boldsymbol{\mu}}|\boldsymbol{\mu}, \boldsymbol{\theta})$ is the probability density function of $w_{\boldsymbol{\mu}}$ under the hypothesis that $\boldsymbol{\mu}$ and $ \boldsymbol{\theta}$ are the true parameters, $w_{\boldsymbol{\mu}, \text{obs}}$ is the observed value of the likelihood ratio, and $F$ is the cumulative distribution of $w_{\boldsymbol{\mu}}$. The boundary of the confidence region for $\boldsymbol{\mu}$, with confidence level $1-\alpha$, is found from the $p$-value of Eq.~\eqref{eq:pvalue} by solving $p_{\boldsymbol{\mu}} = \alpha$.  This gives a region in parameter space that satisfies
\begin{equation}
    \text{Prob}(\boldsymbol{\mu} \in \text{confidence region})\geq 1-\alpha\,.
\end{equation}

When the model has a single parameter of interest $\mu$, it is possible to define another test statistic called the likelihood root as the square root of the likelihood ratio, multiplied by the sign of $\hat{\mu}-\mu$:
\begin{equation}
\label{eq:lik_root}
    r_\mu = \text{sign}(\hat{\mu}-\mu)\sqrt{2\left[\ell(\hat{\mu}, \hat{\boldsymbol{\theta}}) - \ell(\mu, \doublehat{\boldsymbol{\theta}})\right]}\,.
\end{equation}
In contrast to the likelihood ratio, the likelihood root can be defined only when there is a single parameter of interest. However, while the likelihood ratio gives a two-sided test, the likelihood root allows for one-sided tests as well.  The statistic $r_\mu$ can be used to compute $p$-values in the same way this is done for the likelihood ratio using its density function $f(r_\mu|\mu, \boldsymbol{\theta})$.


In many realistic applications, finding the probability density functions $f(w_{\boldsymbol{\mu}}|\boldsymbol{\mu}, \boldsymbol{\theta})$ or $f(r_\mu|\mu, \boldsymbol{\theta})$ is a major challenge since they are usually not known in closed form. Monte Carlo simulations are often used to compute them, but this can be very time-consuming for complex models with many measurements.

It is possible, however, to avoid the numerical computation of $f(w_{\boldsymbol{\mu}})$ and $f(r_\mu)$ in the \textit{asymptotic limit}, in which all the MLEs of the model are Gaussian distributed. This limit is typically reached when the experimental sample size $n$ approaches infinity, i.e., in the so-called \textit{large sample limit}, where the MLEs have a Gaussian distribution with an error term of order $\mathcal{O}(n^{-1/2})$. If this condition holds, $w_{\boldsymbol{\mu}}$ follows a chi-square distribution with $M$ degrees of freedom ($\chi_M^2$),
\begin{equation}
\label{eq:LR_chi2}
w_{\boldsymbol{\mu}} \sim \chi^2_{M} + \mathcal{O}(n^{-1})\,,
\end{equation}
where $M$ is the number of parameters of interest.  In contrast, in the asymptotic limit $r_\mu$ follows a normal distribution with mean $0$ and standard deviation $1$:
\begin{equation}
\label{eq:R_stdNorm}
    r_\mu \sim \mathcal{N}(0,1) + \mathcal{O}(n^{-1/2})\,.
\end{equation}
As already noted, $n$ generally represents the sample size of the experiment, but it can also be another parameter of the likelihood that controls the convergence of the likelihood to the asymptotic limit. It is important to note that the asymptotic distribution of $r_\mu$ exhibits a larger error term compared to that of $w_\mu$. This behavior reflects a trade-off for the flexibility of performing one-sided tests.

In the asymptotic limit, one can show that the profile log-likelihood is approximated by
\begin{equation}
\label{eq:logLik_approx}
    \ell(\boldsymbol{\mu}, \doublehat{\boldsymbol{\theta}}) \simeq \ell(\hat{\boldsymbol{\mu}}, \hat{\boldsymbol{\theta}}) - \frac{1}{2}(\hat{\boldsymbol{\mu}}-\boldsymbol{\mu})V^{-1}(\hat{\boldsymbol{\mu}}-\boldsymbol{\mu})\,,
\end{equation}
where $V_{ij}=\text{cov}[\hat{\mu}_i,\hat{\mu}_j]$.  This can be found using the \textit{observed information matrix} $j_{ij}(\boldsymbol{\hat{\boldsymbol{\psi}}})$,
\begin{equation}
\label{eq:information_matrix}
    j_{ij}(\boldsymbol{\hat{\psi}})=-\frac{\partial^2\ell}{\partial \psi_i\partial \psi_j}\Bigr|_{\boldsymbol{\hat{\psi}}}\,,
\end{equation}
\noindent where $\boldsymbol{\psi} = (\boldsymbol{\mu}, \boldsymbol{\theta})$ represents all of the parameters.  The inverse of the matrix $j$ gives the covariance of all the estimators, $U_{ij} = \mbox{cov}[\hat{\psi}_i, \hat{\psi}_j] = j^{-1}(\hat{\boldsymbol{\psi}}_{ij})$, from which the submatrix $V_{ij}=\text{cov}[\hat{\mu}_i,\hat{\mu}_j]$ can be extracted. 
Under assumption of these approximations, the likelihood ratio is
given by
\begin{equation}
\label{eq:LR_approx}
    w_{\boldsymbol{\mu}}= (\hat{\boldsymbol{\mu}}-\boldsymbol{\mu})V^{-1}(\hat{\boldsymbol{\mu}}-\boldsymbol{\mu})\,.
\end{equation}
Deviations from the quadratic approximations of the likelihood root and the profile likelihood are expected when the conditions of the asymptotic limit are not satisfied.

\section{Higher-order asymptotic corrections}
\label{sec:higher_order_asymp}

When the MLEs of the model parameters are not Gaussian distributed, which usually happens when the experimental sample size is small, the distributions of the statistics $w_{\boldsymbol{\mu}}$ and $r_{\mu}$ deviate from their asymptotic forms.  In such instances, there are two potential strategies to derive test statistics with known distributions: one approach is to refine the approximation of the test statistics' distributions, and the other is to modify the test statistics themselves such that their distributions are more accurately approximated by the asymptotic formulae, even for small sample sizes. An example of the former is given in Ref.~\cite{bib:Xia2021}; in this paper, we focus on the latter approach. 

Specifically, this section explores two potential solutions, the $r^\ast$ approximation~\cite{bib:Barndorff-Nielsen1980,bib:Barndorff-Nielsen1983,bib:Barndorff-Nielsen1986,bib:Barndorff-Nielsen1990} and the Bartlett correction~\cite{bib:Bartlett1937,bib:Lawley1956}. The aim is to derive corrections for $w_\mu$ and $r_\mu$ such that the distributions of the refined statistics, denoted by $w^\ast_\mu$ and $r^\ast_\mu$, can be more precisely approximated by the asymptotic distributions outlined earlier, with error terms of order $\mathcal{O}(n^{-3/2})$ or smaller. Moreover, these enhanced statistics are constructed such that P$(S^\ast > S_{\rm obs}^\ast) = P(S > S_{\rm obs})$ + $\mathcal{O}(n^{-1/2})$, where $S$ represents one of the original statistics and $S^\ast$ corresponds to its improved, higher-order counterpart. This condition ensures that $p$-values computed using the improved statistics are equivalent to those computed with the original ones, up to error terms of order $n^{-1/2}$ or smaller.

\subsection{The $r^*$ approximation}
\label{sec:p_star_aprox}

The asymptotic distributions of the likelihood root and the likelihood ratio are derived from the assumption that the MLEs are Gaussian in the asymptotic limit. But this assumption is only valid up to error terms of order $\mathcal{O}(n^{-1/2})$. A major development in likelihood-based inference has been to improve the approximation of the distributions of the MLEs. For models with a single parameter $\mu$, the Barndorff-Nielsen $p^\ast$ approximation 
\cite{bib:Barndorff-Nielsen1980,bib:Barndorff-Nielsen1983,bib:Barndorff-Nielsen1986,bib:Barndorff-Nielsen1990} is the basic higher-order approximation to the distribution of $\hat{\mu}$:
\begin{equation}
\label{eq:pstar_def}
    f(\hat{\mu}) \simeq p^\ast(\hat{\mu}) \equiv c\,|j(\hat{\mu})|^{1/2}\,e^{-w_\mu/2}\,.
\end{equation}
Here $w_{\mu}$ is the likelihood ratio as defined in Eq.~\eqref{eq:lik_ratio}, $c$ is a normalization constant equal to $1/\sqrt{2\pi}(1+\mathcal{O}(n^{-1}))$, and $j = -\frac{\partial^2 \ell}{\partial\mu^2}$ represents the observed information. Notably, the error term on the $p^\ast$ approximation is of order $n^{-3/2}$. This represents a significant improvement when compared to the $\mathcal{O}(n^{-1/2})$ error term associated with the first-order Gaussian approximation. 
 
When the $p^\ast$ approximation is expanded to $\mathcal{O}(n^{-1/2})$ the Gaussian density of $\hat{\mu}$ is recovered. At this order, the likelihood ratio $w_{\mu}$ can be approximated using Eq.~\eqref{eq:LR_approx} as
\begin{equation}
    w_\mu = (\hat{\mu}-\mu)^2|j(\hat{\mu})| + \mathcal{O}_p(n^{-1/2})\,,
\end{equation}
where $\mathcal{O}_p$ denotes convergence in probability. In this way the $p^{\ast}$ approximation reduces to a Gaussian with mean $\mu$ and standard deviation $|j(\hat{\mu})|$:  
\begin{equation}
    f(\hat{\mu})=\frac{1}{\sqrt{2\pi}}|j(\hat{\mu})|^{1/2}e^{-\frac{(\hat{\mu}-\mu)^2}{2|j(\hat{\mu})|^{-1}}}+\mathcal{O}(n^{-1/2})\,,
\end{equation}
where $j$ is a constant in $\mu$ at order $n^{-1/2}$.

The $p^\ast$ approximation provides a way to modify the statistic $r_\mu$ to reduce the error on its asymptotic distribution. In particular, through an integration of Eq.~\eqref{eq:pstar_def} (see \cite{bib:Barndorff-Nielsen1986}), it can be proved that the modified statistic 
\begin{equation}
\label{eq:Rstar_def}
    r_\mu^\ast = r_\mu + \frac{1}{r_\mu}\log\frac{q_\mu}{r_\mu}\,
\end{equation}
follows a  standard normal distribution with an error term of order $n^{-3/2}$. Equation~\eqref{eq:Rstar_def} for $r_\mu^\ast$ involves a correction term $q_\mu$, whose exact definition is given later in this section. As the model approaches the asymptotic limit, this correction term converges towards $r_\mu$, thereby leading $r^\ast_\mu$ to approach $r_\mu$. Conversely, when the model diverges from the asymptotic limit, this correction term modifies $r_\mu$ so that the asymptotic distribution of $r_\mu^\ast$ has a reduced error term:
\begin{equation}
    r_\mu^\ast \sim \mathcal{N}(0,1) + \mathcal{O}(n^{-3/2})\,.
\end{equation}
That is, the error on the asymptotic distribution of $r_\mu^\ast$ falls off three powers of $n^{-1/2}$ faster than that of the likelihood root (see Eq.~\eqref{eq:lik_root}). In addition, by squaring $r_\mu^\ast$ one obtains the statistic $(r_\mu^{\ast})^2$, which is asymptotically distributed as chi-squared with one degree of freedom. This can be interpreted as a higher-order correction to the likelihood ratio statistic $w_\mu$.

An intuitive interpretation of the $r_\mu^\ast$ statistic can be obtained, despite its non-trivial derivation (see, e.g., \cite{bib:Brazzale2007}). One can show that the new $r_\mu^\ast$ statistic is related to $r_\mu$ by
\begin{equation}
    r_\mu^\ast = \frac{r_\mu - \text{E}[r_\mu]}{\text{V}[r_\mu]^{1/2}} + \mathcal{O}_p(n^{-3/2})\,,
\end{equation}
where $\text{E}[r_\mu]$ and $\text{V}[r_\mu]$ are the expectation value and variance of $r_\mu$. This equation says that, up to errors of order $n^{-3/2}$, $r_\mu^\ast$ represents the standardized version of $r_\mu$. Furthermore, this equation provides a method to approximately compute $r_\mu^\ast$ using MC to estimate $\text{E}[r_\mu]$ and $\text{V}[r_\mu]$ as an alternative to the analytical computation of $r^\ast_\mu$.

The analytical computation of $r^\ast_\mu$ requires the correction term $q_{\mu}$, which can be found in different ways depending on the characteristics of the statistical model. Details can be found in \cite{bib:Brazzale2007}; here we summarize the main results. For models with one parameter of interest and no nuisance parameters, $q_\mu$ can be found as
\begin{equation}
\label{eq:q_def}
    q_\mu = \left(\frac{\partial \ell}{\partial \hat{\mu}}\Bigr|_{\hat{\mu}}-\frac{\partial \ell}{\partial \hat{\mu}}\Bigr|_{\mu}\right)j(\hat{\mu})^{1/2}\,,
\end{equation}
where the first derivative is computed for $\mu=\hat{\mu}$ and the second for the value of $\mu$ being tested. For statistical models that include nuisance parameters, if we assume that the full parameter space can be written as $\boldsymbol{\psi}=(\mu, \boldsymbol{\theta})$, where $\mu$ is the parameter of interest and $\boldsymbol{\theta}$ is a vector of $N$ nuisance parameters, the correction term $q_\mu$ can be found from
\begin{equation}
\small
\label{eq:q1_def}
    q_{\mu} = \frac{\text{det}\left[\ell_{\hat{\boldsymbol{\psi}}}(\hat{\mu}, \hat{\boldsymbol{\theta}})-\ell_{\hat{\boldsymbol{\psi}}}(\mu, \hat{\hat{\boldsymbol{\theta}}}),\quad\ell_{\boldsymbol{\theta}\hat{\boldsymbol{\psi}}}(\mu, \hat{\hat{\boldsymbol{\theta}}})\right]}{\text{det}\left[\ell_{\boldsymbol{\psi}\hat{\boldsymbol{\psi}}}(\hat{\mu}, \hat{\boldsymbol{\theta}})\right]} \left( \frac{\text{det}[j_{\boldsymbol{\psi}\boldsymbol{\psi}}(\hat{\mu}, \hat{\boldsymbol{\theta}})]}{\text{det}[j_{\boldsymbol{\theta}\boldsymbol{\theta}}(\mu, \hat{\hat{\boldsymbol{\theta}}})]}\right)^{1/2}\,,
\end{equation}
where $j$ is the information matrix
\begin{equation}
    j_{\boldsymbol{\psi}\boldsymbol{\psi}}(\boldsymbol{\psi}) = -\frac{\partial^2\ell(\boldsymbol{\psi})}{\partial\boldsymbol{\psi}\partial\boldsymbol{\psi}^T}\,.
\end{equation}
The subscripts on $\ell$ indicate derivatives; for example, $\ell_{\boldsymbol{\psi}}$ is the gradient of $\ell$ with respect to the parameters $\boldsymbol{\psi}$. The numerator and denominator of Eq.~\eqref{eq:q1_def} contain determinants of $(N+1)\times (N+1)$ matrices, where $N+1$ is the dimension of the full parameter space. The matrix in the numerator of the first factor has the $(N+1)$-dimensional vector $\ell_{\hat{\boldsymbol{\psi}}}(\hat{\mu}, \hat{\boldsymbol{\theta}})-\ell_{\hat{\boldsymbol{\psi}}}(\mu, \hat{\hat{\boldsymbol{\theta}}})$ as its first column, while the remaining $(N+1)\times N$ section of the matrix is given by $\ell_{\boldsymbol{\theta}\hat{\boldsymbol{\psi}}}(\mu, \hat{\hat{\boldsymbol{\theta}}})$, i.e. the matrix of the second derivatives of $\ell$ with respect to the nuisance parameters $\boldsymbol{\theta}$ as the first index, and with respect to the full parameters vector $\boldsymbol{\psi}$ as the second index.

Equations \eqref{eq:q_def} and \eqref{eq:q1_def} contain derivatives of the likelihood with respect to the MLEs; therefore they require the observed data $\boldsymbol{y}$, or equivalently the likelihood, to be expressed as an explicit function of them, which is not always possible. In such cases, one can replace $p^*$ by an alternative quantity $p_{\rm TEM}$, where TEM stands for Tangent Exponential Model (see \cite{bib:Davison2022}). This expression for the density function of $\hat{\mu}$ exploits a local approximation of the likelihood with an exponential family model. The \textit{canonical parameters} $\phi$ of the approximating model are defined by
\begin{equation}
    \label{eq:canonical_params_def}
    \boldsymbol{\phi}^T (\boldsymbol{\psi}, \boldsymbol{y_{\rm obs}}) = \sum_{i=1}^{n} \frac{\partial \ell }{\partial y_i}\Bigr|_{\boldsymbol{y_{\rm obs}}} \,\times \, V\,,
\end{equation}
and they can be used to derive an alternative expression for the correction term $q_\mu$. Here $\boldsymbol{y_{\rm obs}}$ is the $n$-dimensional vector of the observed data and $V$ is an $n \times (N+1)$ matrix defined as
\begin{equation}
    \label{eq:Vmatrix_def}
    V = - \left(\frac{\partial\boldsymbol{z}}{\partial \boldsymbol{y}^T}\right)^{-1}\left(\frac{\partial\boldsymbol{z}}{\partial \boldsymbol{\boldsymbol{\psi}}^T}\right)\Bigr|_{\hat{\boldsymbol{\psi}}_{\rm obs}}\,.
\end{equation}
In the last expression, $\boldsymbol{z} = (z_1(y_1),..., z_N(y_n))$ is a vector of \textit{pivotal quantities}. Pivotal quantities are transformations of data that have a fixed distribution under the model, i.e., their distribution does not depend on the parameters of the model. Such a vector always exists in the form of cumulative distributions $F(y_i)$, which are always uniformly distributed in $[0,1]$ for continuous data. But alternative choices are often available, e.g., for a Gaussian distributed random variable $y$ with mean $\mu$ and standard deviation $\sigma$ one 
can define the pivotal statistic $z = (y - \mu)/\sigma$, whose distribution is a standard normal for any chosen $\mu$ and $\sigma$. 

Using the canonical parameters $\boldsymbol{\phi}$, alternative equations for the correction term $q_\mu$ can be derived. For models with one parameter of interest and no nuisance parameters the formula for $q_\mu$ becomes
\begin{equation}
\label{eq:q_def_2}
    q_\mu = \left[\phi(\hat{\mu})-\phi(\mu)\right]j(\hat{\mu})^{1/2}\left|\frac{\partial \phi}{\partial \mu}(\hat{\mu})\right|^{-1}\,.
\end{equation}
Whereas, for  models that include nuisance parameters, $q_\mu$ is given by
\begin{equation}
\label{eq:q2_def}
\small
    q_{\mu} = \frac{\text{det}\left[\boldsymbol{\phi}(\hat{\mu}, \hat{\boldsymbol{\theta}})-\boldsymbol{\phi}(\mu, \hat{\hat{\boldsymbol{\theta}}}),\quad\boldsymbol{\phi}_{\boldsymbol{\theta}}(\mu, \hat{\hat{\boldsymbol{\theta}}})\right]}{\text{det}\left[\boldsymbol{\phi}_{\boldsymbol{\psi}}(\hat{\mu}, \hat{\boldsymbol{\theta}})\right]}\left( \frac{\text{det}[j_{\boldsymbol{\psi}\boldsymbol{\psi}}(\hat{\mu}, \hat{\boldsymbol{\theta}})]}{\text{det}[j_{\boldsymbol{\theta}\boldsymbol{\theta}}(\mu, \hat{\hat{\boldsymbol{\theta}}})]}\right)^{1/2}\,.
\end{equation}
The definition of $V$ given by Eq.~\eqref{eq:Vmatrix_def} applies to the case where $y_i$ are continuous variables. If instead they are discrete, $V$ can be obtained from
\begin{equation}
    \label{eq:Vmatrix_discrete_def}
    V = \frac{\text{dE}[\boldsymbol{y}|\boldsymbol{\psi}]}{d\boldsymbol{\psi}^T}\Bigr|_{{\boldsymbol{\psi}}={\hat{\boldsymbol{\psi}}}}\,.
\end{equation}
In addition, in the definition of the canonical parameters $\boldsymbol{\phi}$ given by Eq.~\eqref{eq:canonical_params_def} the derivatives $\partial \ell / \partial y_i$ are replaced by $\partial \log(f_i)/\partial y_i$, where $f_i(y_i)$ is the probability distribution of $y_i$.
More details on how to compute $r^\ast_\mu$ for applications involving discrete data can be found in \cite{bib:Davison2006}.

\subsection{The Bartlett correction}
\label{sec:bartlett_correction}

A different approach to higher-order asymptotics due to \\Bartlett \cite{bib:Bartlett1937} involves a scaling of the likelihood ratio statistic, rather than a correction to the distributions of the MLEs. Bartlett's argument is as follows. For a model with $M$ parameters of interest $\boldsymbol{\mu}$ the likelihood ratio follows a chi-square distribution with $M$ degrees of freedom ($\chi^2_M$) and its expectation value is
\begin{equation}
    \text{E}[w_{\boldsymbol{\mu}}] = M + b \,,
\end{equation}
where $b$ is the correction to the asymptotic expectation value. The modified statistic
\begin{equation}
\label{eq:wstarDef}
    w_{\boldsymbol{\mu}}^\ast = w_{\boldsymbol{\mu}}\, \frac{M}{\text{E}[w_{\mu}]} \equiv  \frac{w_{\boldsymbol{\mu}}}{1+b/M}
\end{equation}
follows a distribution closer to the asymptotic $\chi_M^2$.  The quantity $b = \text{E}[w_{\boldsymbol{\mu}}] - M$ characterizes the size of the Bartlett correction. 

In many realistic applications, $b$ cannot be computed exactly. In such scenarios, two possible approaches exist. The first is to estimate $\text{E}[w_{\boldsymbol{\mu}}]$ using MC methods, while the second is to approximate it perturbatively using a result provided by Lawley.

Lawley~\cite{bib:Lawley1956} developed a general method to compute the expectation value up to $\mathcal{O}(n^{-2})$ proving that all the cumulants of $w_{\boldsymbol{\mu}}^\ast$ match with the cumulants of a $\chi_M^2$ distribution up to this order. Specifically, Lawley's formula is based on a quartic expansion of both the likelihood ratio and the score equation, $\frac{\partial \ell }{\partial \boldsymbol{\mu}}(\boldsymbol{\hat{\mu}})=0$, in powers of $\hat{\mu}_i-\mu_i$ (see, e.g., \cite{bib:Cordeiro2014}), where $i$ is an index running over the parameter space of the model. The two expansions can be combined to obtain an approximation of the expectation value
\begin{equation}
    \text{E}[w_{\boldsymbol{\mu}}] = 2\text{E}[\ell(\hat{\boldsymbol{\mu}})-\ell(\boldsymbol{\mu})] = M + \epsilon_M + \mathcal{O}(n^{-2})\,,
\end{equation}
where $\epsilon_M$ here represents the Bartlett correction factor $b$ computed to order $n^{-1}$ using the Lawley method. The correction term $\epsilon_M$ has a complicated structure involving derivatives of the likelihood up to the fourth order and their expectation values. Nevertheless, for many applications, it is possible to compute it analytically. Specifically, for a model with $M$ parameters of interest and without any nuisance parameters (the case with nuisance parameters will be considered later in this section) $\epsilon_M$ can be written as
\begin{equation}
\label{eq:lawely_correction_def}
    \epsilon_M = \sum_{rstu} \lambda_{rstu} - \sum_{rstuvw} \lambda_{rstuvw}\,,
\end{equation}
where the indexes $r$,$s$,$t$,$u$,$v$, and $w$ label all the $M$ parameters of the model (see, e.g., Ref.~\cite{bib:Cordeiro2014}). The two terms inside the sum are
\begin{equation}
\label{eq:lawley_terms_def}
\begin{aligned}
    &\lambda_{rstu} = k^{rs}k^{tu}\left(\frac{1}{4}k_{rstu}-k_{rst}^{(u)}+k_{rs}^{(tu)}\right)\,, \\ 
    &\lambda_{rstuvw} =  k^{rs}k^{tu}k^{vw} \Bigg(\frac{1}{6}k_{rtv}k_{suw}+\frac{1}{4}k_{rtu}k_{svw}-k_{rtv}k_{sw}^{(u)}\\
    &\phantom{=} -k_{rtu}k_{sw}^{(v)}+k_{rt}^{(v)}k_{sw}^{(v)}+k_{rt}^{(u)}k_{sw}^{(v)} \Bigg) \,.
\end{aligned}
\end{equation}
The terms inside the above definitions can be computed as 
\begin{equation}
\label{eq:k-terms1_def}
\begin{split}
&k_{rs} = E\left[\frac{\partial^2 \ell}{\partial\mu_r\partial \mu_s}\right]\,,\\
&k_{rst} = E\left[\frac{\partial^3\ell}{\partial\mu_r\partial\mu_s\partial\mu_t}\right]\,,\\
&k_{rstu} = E\left[\frac{\partial^4\ell}{\partial\mu_r\partial\mu_s\partial\mu_t\partial\mu_u}\right]\,,
\end{split}
\end{equation}
and
\begin{equation}
\begin{split}
\label{eq:k-terms2_def}
&k_{rs}^{(t)} =\frac{\partial k_{rs}}{\partial\mu_t}\,,\\
&k_{rs}^{(tu)} =\frac{\partial^2 k_{rs}}{\partial\mu_t\partial \mu_u}\,,\\
&k_{rst}^{(u)} =\frac{\partial k_{rst}}{\partial\mu_u}\,,
\end{split}
\end{equation}
where the matrices with upper indices are the inverses of the corresponding matrices with lower indices. The general expression for the Bartlett correction is quite involved, but its computation is not conceptually complicated, as it only involves computing derivatives and expectation values of them.

Often the parameters can be split into two subsets: parameters of interest $\boldsymbol{\mu} = (\mu_1,\,...,\, \mu_M)$, and nuisance parameters $\boldsymbol{\theta} = (\theta_1, \,..., \, \theta_{N})$, and one is typically interested in testing specific parameter values in $\boldsymbol{\mu}$ space. In such scenarios, the Lawley formula to compute the expectation of $w_{\boldsymbol{\mu}}$ is given by
\begin{equation}
\label{eq:lawelyComposite_correction_def}
\begin{split}
    \text{E}[w_{\boldsymbol{\mu}}] &= 2\text{E}[\ell(\boldsymbol{\hat{\mu}},\hat{\boldsymbol{\theta}})-\ell(\boldsymbol{\mu},\hat{\hat{\boldsymbol{\theta}}})]\\*[0.2 cm]
    &= 2\text{E}[\ell(\boldsymbol{\hat{\mu}},\hat{\boldsymbol{\theta}})-\ell(\boldsymbol{\mu},\boldsymbol{\theta})] - 2\text{E}[\ell(\boldsymbol{\mu},\hat{\hat{\boldsymbol{\theta}}})-\ell(\boldsymbol{\mu},\boldsymbol{\theta})]\\*[0.2 cm] 
    &= M + N +\epsilon_{N+M} - N - \epsilon_{N} + \mathcal{O}(n^{-2})\\*[0.2 cm] 
    &= M + \epsilon_{N+M} - \epsilon_{N} + \mathcal{O}(n^{-2})\,.
\end{split}
\end{equation}
The notation $\epsilon_{N+M}$ indicates that the summation in Eq.\eqref{eq:lawely_correction_def} is performed over all the indices labeling the full parameters space, wheres the notation $\epsilon_{N}$ indicates that the summation is only performed over indices of the nuisance parameters. However, a more efficient way to compute Eq.~\eqref{eq:lawelyComposite_correction_def}, is to directly calculate the difference $\epsilon_{N+M} - \epsilon_{N}$ by summing the terms in Eq.~\eqref{eq:lawley_terms_def} over all permutations of the indices that contain at least one parameter of interest. It is worth noting that, for composite hypotheses, the expectation value of the likelihood ratio is dependent on the nuisance parameters, as the distribution of $w_{\boldsymbol{\mu}}$ still has a dependence on them. Therefore, to evaluate the expectation value, and thus Eq.\eqref{eq:lawelyComposite_correction_def}, one should use $\boldsymbol{\theta}= \hat{\hat{\boldsymbol{\theta}}}({\boldsymbol{\mu}})$, the constrained MLEs of the nuisance parameters.

The Lawley formula \eqref{eq:lawelyComposite_correction_def} is a valuable tool in situations where it is not feasible to compute the exact expectation value of $w_{\boldsymbol{\mu}}$ analytically, a common scenario in realistic applications. An alternative approach is to numerically estimate the expectation value of $w_{\boldsymbol{\mu}}$ using MC methods.  Specifically, E$[w_{\boldsymbol{\mu}}]$ can be estimated by generating data and setting the parameters of interest $\boldsymbol{\mu}$ to the value in the parameter space being tested, and the nuisance parameters $\boldsymbol{\theta}$ to their profiled values $\hat{\hat{\boldsymbol{\theta}}}(\boldsymbol{\mu})$ (also known as parametric bootstrap estimate).

\section{Overview of the Gamma Variance Model}
\label{sec:gammaModel_def}

Having outlined in the preceding section the general formalism for higher-order asymptotic corrections, we now demonstrate their use with
the Gamma Variance Model (GVM) introduced in Ref.~\cite{bib:Cowan2019}. The GVM extends the general likelihood often used in particle physics analysis,
\begin{equation}
\begin{split}
\label{eq:basic_lik}
    L(\boldsymbol{\mu}, \boldsymbol{\theta}) &= P(\boldsymbol{y}|\boldsymbol{\mu}, \boldsymbol{\theta})\times P(\boldsymbol{u}| \boldsymbol{\theta}) \\
    &=  P(\boldsymbol{y}|\boldsymbol{\mu}, \boldsymbol{\theta}) \times \prod\limits_{i=1}^{N}\frac{1}{\sqrt{2\pi\sigma_{u_i}^2}}\
    \exp\left[ -\frac{(u_i-\theta_i)^2}{2\sigma_{u_i}^2} \right] \,,
\end{split}
\end{equation}
where $P(\boldsymbol{y}|\boldsymbol{\mu}, \boldsymbol{\theta})$ denotes the probability density function of the data $\boldsymbol{y}$, which depends on $M$ parameters of interest $\boldsymbol{\mu} = (\mu_1, \ldots, \mu_M)$ and $N$ nuisance parameters $\boldsymbol{\theta} = (\theta_1, \ldots, \theta_N)$.  To provide information on the nuisance parameters one includes $N$ \textit{control measurements} \\ $\boldsymbol{u} = (u_1, \ldots, u_N)$, here assumed to be direct estimates of the nuisance parameters $\boldsymbol{\theta}$ that are independent and Gaussian distributed.  We suppose these are unbiased, (i.e., $\text{E}[\boldsymbol{u}] = \boldsymbol{\theta}$), and their standard deviations $\boldsymbol{\sigma_u}= (\sigma_{u_1},...,\sigma_{u_N})$ are often referred to as \textit{systematic errors}.  The inclusion of nuisance parameters enlarges the model's parameter space, thereby enabling a better approximation of the truth, albeit at the cost of reduced sensitivity to the parameters of interest.  Very often, the values of the standard deviations $\boldsymbol{\sigma_u}$ are assigned by the experimenter and treated as fixed.

The GVM extends this model to address the important situation where the systematic errors are themselves  uncertain by regarding the $\sigma_{u_i}^2$ as adjustable rather than known parameters. The values that one would have assigned to them before are  now treated as independent gamma-distributed estimates $v_i$, i.e.,

\begin{equation}
    v_i \sim \frac{\beta_i^{\alpha_i}}{\Gamma(\alpha_i)}v_i^{\alpha_i-1}e^{-\beta_i v_i}\,.
\end{equation}

\noindent Here the parameters of the gamma distribution $\alpha_i$ and
$\beta_i$ are defined such that the expected value is
$\text{E}[v_i] = \alpha_i / \beta_i$ and the variance is 
$\sigma_{v_i}^2 = \alpha_i / \beta_i^2$.  These are chosen such that $v_i$ is an unbiased estimator for $\sigma_{u_i}^2$, (i.e., $\text{E}[v_i] = \sigma_{u_i}^2$) and the width of the gamma distribution is adjusted 
to reflect the appropriate level of uncertainty by defining

\begin{equation}
\label{eq:eps_def}
    \varepsilon_i \equiv \frac{1}{2}\frac{\sigma_{v_i}}{\text{E}[v_i]}= \frac{1}{2}\frac{\sigma_{v_i}}{\sigma^2_{u_i}}\,,
\end{equation}
which  is a fixed parameter of the model. Using error propagation Eq.~\eqref{eq:eps_def} becomes
\begin{equation}
    \varepsilon_i \simeq \frac{\sigma_{s_i}}{\text{E}[s_i]}\,,
\end{equation}

\noindent where $s_i = \sqrt{v_i}$.  The quantity $\varepsilon_i$ is thus the relative uncertainty on the assigned systematic error (also called a \textit{coefficient of variation}), which we refer to informally as the \textit{relative error-on-error} parameter. Including the $v_i$ as measurements into the likelihood gives
\begin{equation}
\begin{split}
\label{eq:GVM_lik}
        L(\boldsymbol{\mu}, \boldsymbol{\theta}) &=  P(\boldsymbol{y}|\boldsymbol{\mu}, \boldsymbol{\theta}) \\ &\times \prod\limits_{i=1}^{N}\frac{1}{\sqrt{2\pi\sigma_{u_i}^2}}\,e^{-\frac{(u_i-\theta_i)^2}{2\sigma_{u_i}^2}} \frac{\beta_i^{\alpha_i}}{\Gamma(\alpha_i)}v_i^{\alpha_i-1}e^{-\beta_i v_i}\,.
\end{split}
\end{equation}
Although treating the $\sigma^2_{u_i}$ as adjustable in the GVM in effect doubles the number of nuisance parameters in comparison to the model where the $\sigma_{u_i}^2$ are known, one can profile over them in closed form.  After some manipulation (see Ref.~\cite{bib:Cowan2019}), the profile log-likelihood is found to be
\begin{equation}
\label{eq:general_GVM_loglik}
\begin{split}
    &\ell(\boldsymbol{\mu},\boldsymbol{\theta},\boldsymbol{\widehat{\widehat{\sigma^2_{u}}}})=\ell_p(\boldsymbol{\mu},\boldsymbol{\theta})=\log{P(\boldsymbol{y}|\boldsymbol{\mu}, \boldsymbol{\theta})}\\
    &\phantom{{=}} -\frac{1}{2}\sum_{i=1}^N\left(1+\frac{1}{2\varepsilon_i^2}\right)\log{\left[ 1+2\varepsilon_i^2\frac{\left(u_i-\theta_i\right)^2}{v_i}\right]}\,.
\end{split}
\end{equation}

As discussed in Refs.~\cite{bib:Cowan2019}, the Gamma Variance Model leads to interesting and useful consequences for inference about the parameters of interest $\boldsymbol{\mu}$.  In particular, the size of the confidence region for $\boldsymbol{\mu}$ becomes coupled to the goodness of fit, with increasing incompatibility of the input data leading to larger regions.  Furthermore, the point estimate for $\boldsymbol{\mu}$ shows a decreased sensitivity to outliers in the data.  It is therefore of particular interest to apply the GVM in cases where the input values are in tension either among themselves or with the predictions of a hypothesis of interest.   For example, the tension between measured and predicted values of the anomalous muon magnetic moment was explored in Ref.~\cite{bib:Cowan2022}.  The GVM represents a purely frequentist approach to this 
type of problem.  Bayesian methods have been found to yield qualitatively similar results, e.g., in 
Refs.~\cite{bib:Dose2014,bib:Dagostini1999,bib:Cowan2006,bib:Erler2020}.

A practical difficulty with the Gamma Variance Model arises in connection with the use of asymptotic formulae to obtain $p$-values and confidence regions when the $\varepsilon$ parameters exceed a value of around 0.2.   As discussed in Ref.~\cite{bib:Cowan2019}, there is a correspondence between the parameters $\varepsilon_i$ and an effective sample size, $n_i$, which can be found by considering a sample of $n_i$
independent observations of $u_i$ and using their sample variance as an estimate of $\sigma_{u_i}^2$.  This estimator is found to be gamma distributed with an error-on-error parameter $\varepsilon_i$ related to the sample size by
\begin{equation}
    \label{eq:neff_from_r}
    n_i = 1 + \frac{1}{2 \varepsilon_i^2} \;.
\end{equation}

\noindent Thus when $\varepsilon_i$ becomes too large, then $n_i$ drops to become of order unity and the large-sample criterion required for use of asymptotic distributions no longer holds. Values of $\varepsilon$ are expected to be roughly 0.2 to 0.5 or even larger in many applications, which could make it far more difficult to compute $p$-values and confidence regions.  

The breakdown of the asymptotic formulae for large $\varepsilon_i$ can be understood intuitively by expanding the logarithmic term of Eq.~\eqref{eq:general_GVM_loglik} in powers of $\varepsilon_i$:
\begin{equation}
\label{eq:log_expansion}
\begin{split}
    &\left(1+\frac{1}{2\varepsilon_i^2}\right)\log{\left[ 1+2\varepsilon_i^2\frac{\left(u_i-\theta_i\right)^2}{v_i}\right]}\\
    & =\left(1+2\varepsilon_i^2\right)\frac{\left(u_i-\theta_i\right)^2}{v_i}-\varepsilon_i^2\frac{\left(u_i-\theta_i\right)^4}{v_i^2}+\mathcal{O}_p(\varepsilon_i^4)\,.
    \end{split}
\end{equation}
Thus as $\varepsilon_i$ approaches zero, the logarithmic constraint reduces to a quadratic one, associated with a Gaussian constraint for the nuisance parameter, leading to the asymptotic distributions for the statistics $w_{\mu}$ and $r_{\mu}$ discussed above.  However, for large $\varepsilon_i$, the Gamma Variance Model deviates from the quadratic approximation by an error term that begins at $\mathcal{O}_p(\varepsilon_i^2)$, as can be seen in Eq.~\eqref{eq:log_expansion}. 

Consequently, when $\varepsilon_i$ is not equal to zero, the asymptotic formulae used to obtain $p$-values and confidence regions are not guaranteed to represent valid approximations.  Furthermore, the interval of convergence of the logarithm in Eq.~\eqref{eq:general_GVM_loglik} is
\begin{equation}
\label{eq:conv_radius}
    2\varepsilon_i^2\frac{\left(u_i-\theta_i\right)^2}{v_i}<1\,,
\end{equation}
and thus the asymptotic formulae are not expected to give accurate approximations if the above condition is not satisfied. 

In principle, this difficulty can be overcome by using Monte Carlo calculations, but this can entail substantial additional work and computing time.  It is therefore valuable to have a method of finding $p$-values and confidence regions without MC, and thus the primary goal of this paper is to investigate the use of higher-order asymptotics with the GVM to obtain results that remain accurate even for large $\varepsilon_i$.  

\section{Single-measurement model}
\label{sec:smm}

In order to investigate the asymptotic properties of a statistical model with uncertain error parameters, it is convenient to use
the simple model introduced in Ref.~\cite{bib:Cowan2019}. Here for completeness we reproduce several results shown in that paper using the Bartlett correction and extend them in Sec.~\ref{sec:hoasmm} using the $r^*$ approximation.

The single-measurement model  describes a single Gaussian distributed measurement $y$ with mean $\mu$ and standard deviation $\sigma$. We take $\mu$ to be the parameter of interest and $\sigma^2$ to be a nuisance parameter constrained by an independent gamma-distributed estimate $v$. Therefore, the likelihood is
\begin{equation}
\label{eq:single_mes_model}
    L(\mu, \sigma^2) =\frac{1}{\sqrt{2\pi\sigma^2}}e^{-(y-\mu)^2/2\sigma^2}\frac{\beta}{\Gamma(\alpha)}v^{\alpha-1}e^{-\beta v}\,,
\end{equation}
where $\alpha=1/4\varepsilon^2$ and $\beta=1/(4\varepsilon^2\sigma^2)$, and $\varepsilon$ is the relative error on the standard deviation $\sigma$.  
The log-likelihood of the model is given by
\begin{equation}
    \ell(\mu, \sigma^2)=-\frac{1}{2}\frac{(y-\mu)^2}{\sigma^2}-\left(\frac{1}{2}+\frac{1}{4\varepsilon^2}\right)\log{\sigma^2}-\frac{v}{4\varepsilon^2\sigma^2}\,.
\end{equation}
The goal is to compute the likelihood ratio $w_{\mu}$ (see Eq.~\eqref{eq:lik_ratio}) to study its asymptotic properties and to apply to it the higher-order corrections defined in Sec.\ref{sec:higher_order_asymp}. This requires the estimators
\begin{equation}
\label{eq:inMes_MLE}
\begin{split}
    &\hat{\mu}=y\,,\\
    &\widehat{\sigma^2}=\frac{v}{1+2\varepsilon^2}\,,\\
    &\widehat{\widehat{\sigma^2}}=\frac{v+2\varepsilon^2(y-\mu)^2}{1+2\varepsilon^2}\,.
\end{split}
\end{equation}
With the help of the above expressions it is easy to derive the likelihood ratio
$w_{\mu}$,
\begin{equation}
    w_\mu=\left(1+\frac{1}{2\varepsilon^2}\right)\log{\left[1+2\varepsilon^2\frac{(y-\mu)^2}{v}\right]}\,,
\end{equation}
which, in the limit $\varepsilon \rightarrow 0$, becomes
\begin{equation}
    w_\mu=\frac{(y-\mu)^2}{v}+\mathcal{O}_p(\varepsilon^2)\,.
\end{equation}
In this limit, the likelihood ratio can be approximated by a quadratic expression, as expected in the asymptotic limit. As seen previously, the parameter $\varepsilon$ is related to an effective sample size, as it measures the extent to which the model deviates from the asymptotic limit. In particular, it is expected that the distribution of $w_{\mu}$ should deviate from its asymptotic $\chi_1^2$ distribution by an error term of order $\mathcal{O}(\varepsilon^2)$. 

\begin{figure*}[t]
    \begin{center}
    \includegraphics[width=0.49\textwidth]{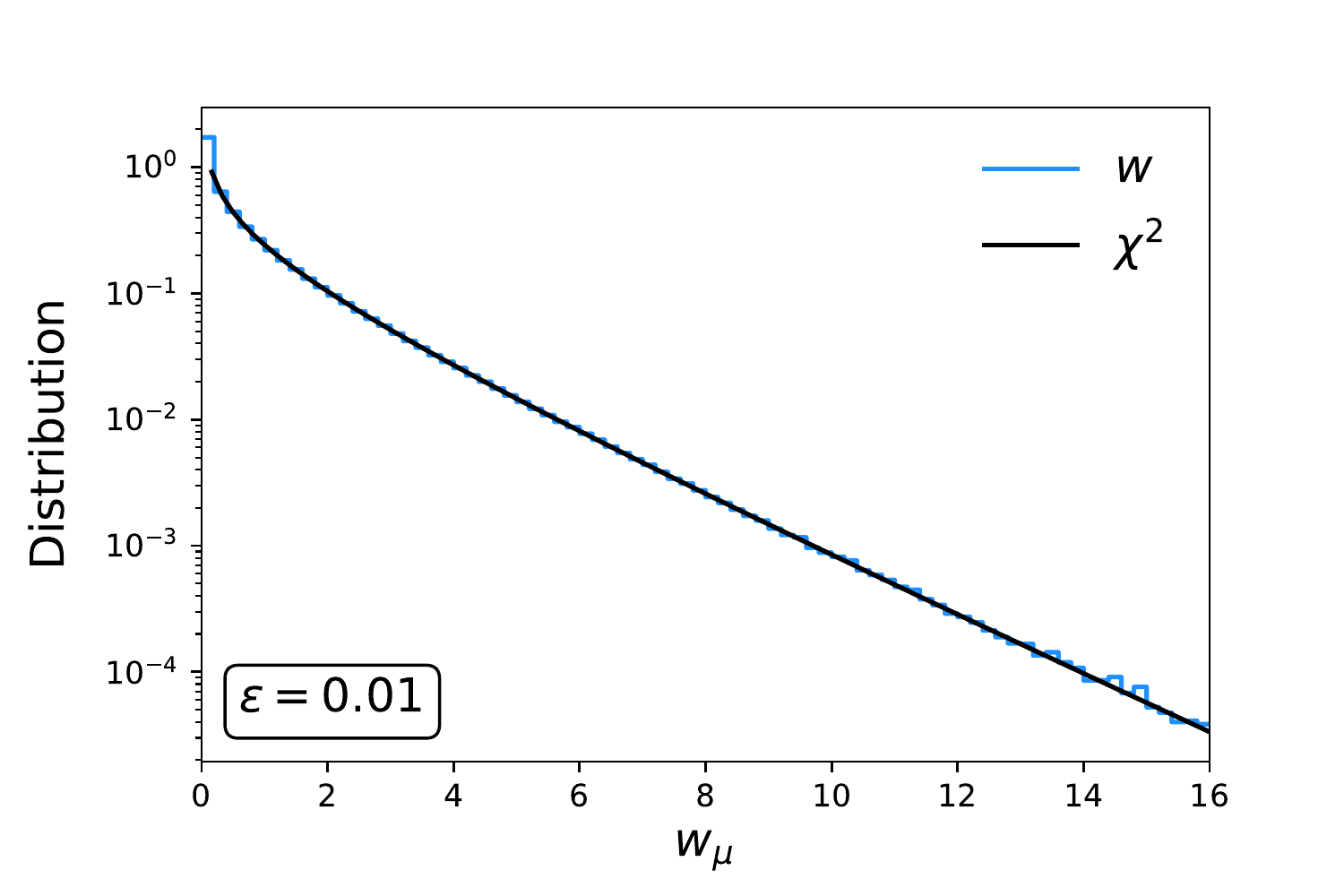}
    \includegraphics[width=0.49\textwidth]{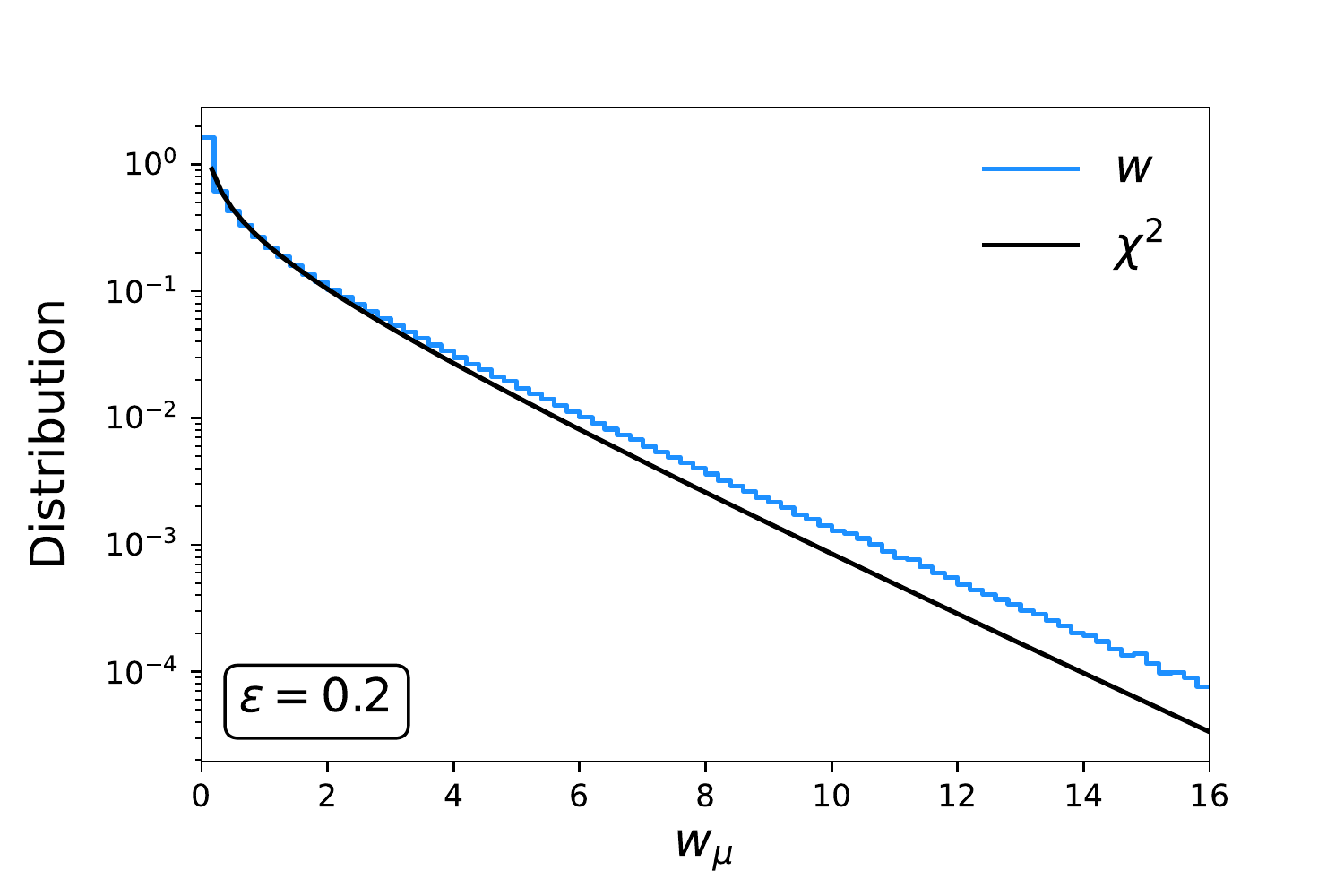}
    \includegraphics[width=0.49\textwidth]{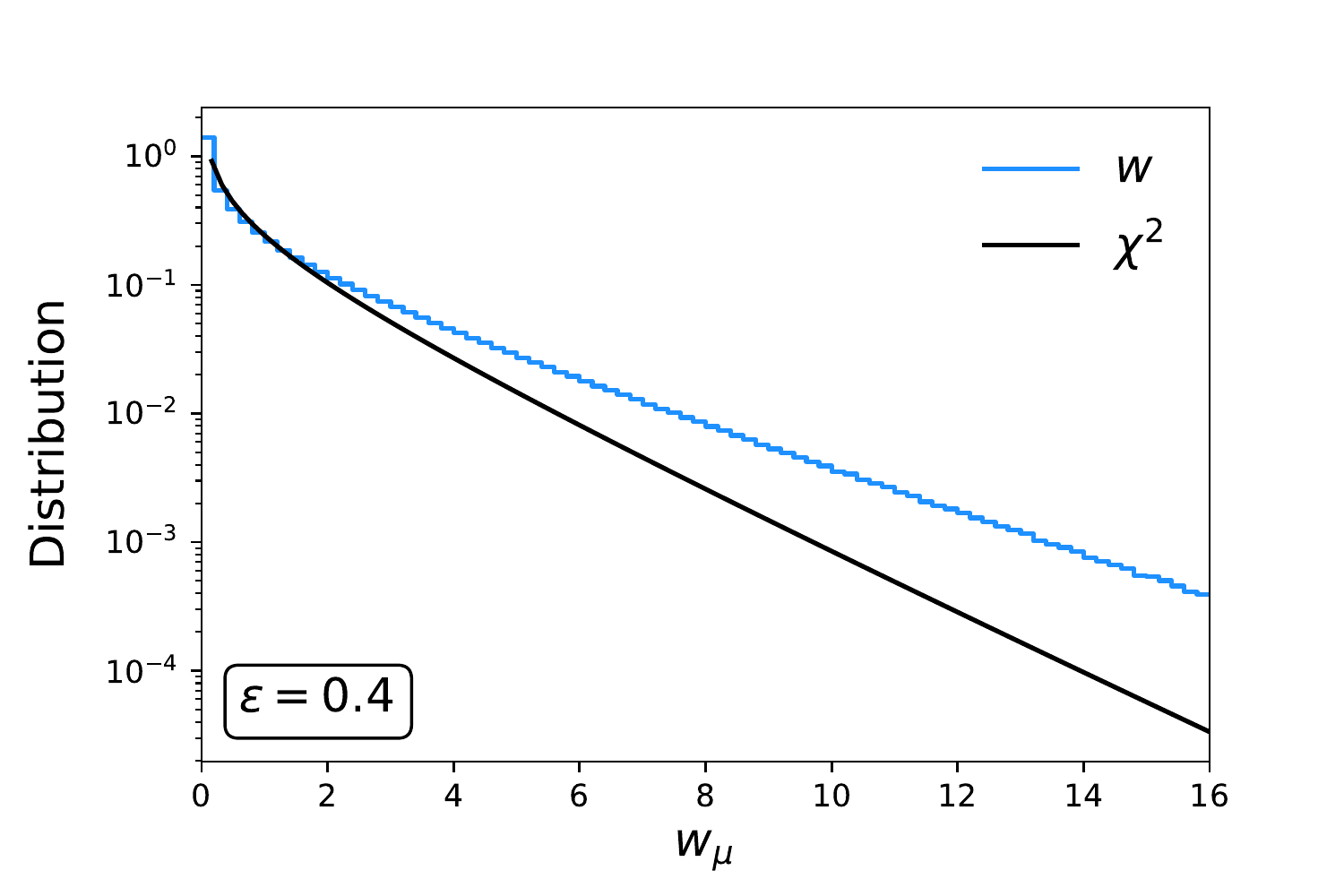}
    \includegraphics[width=0.49\textwidth]{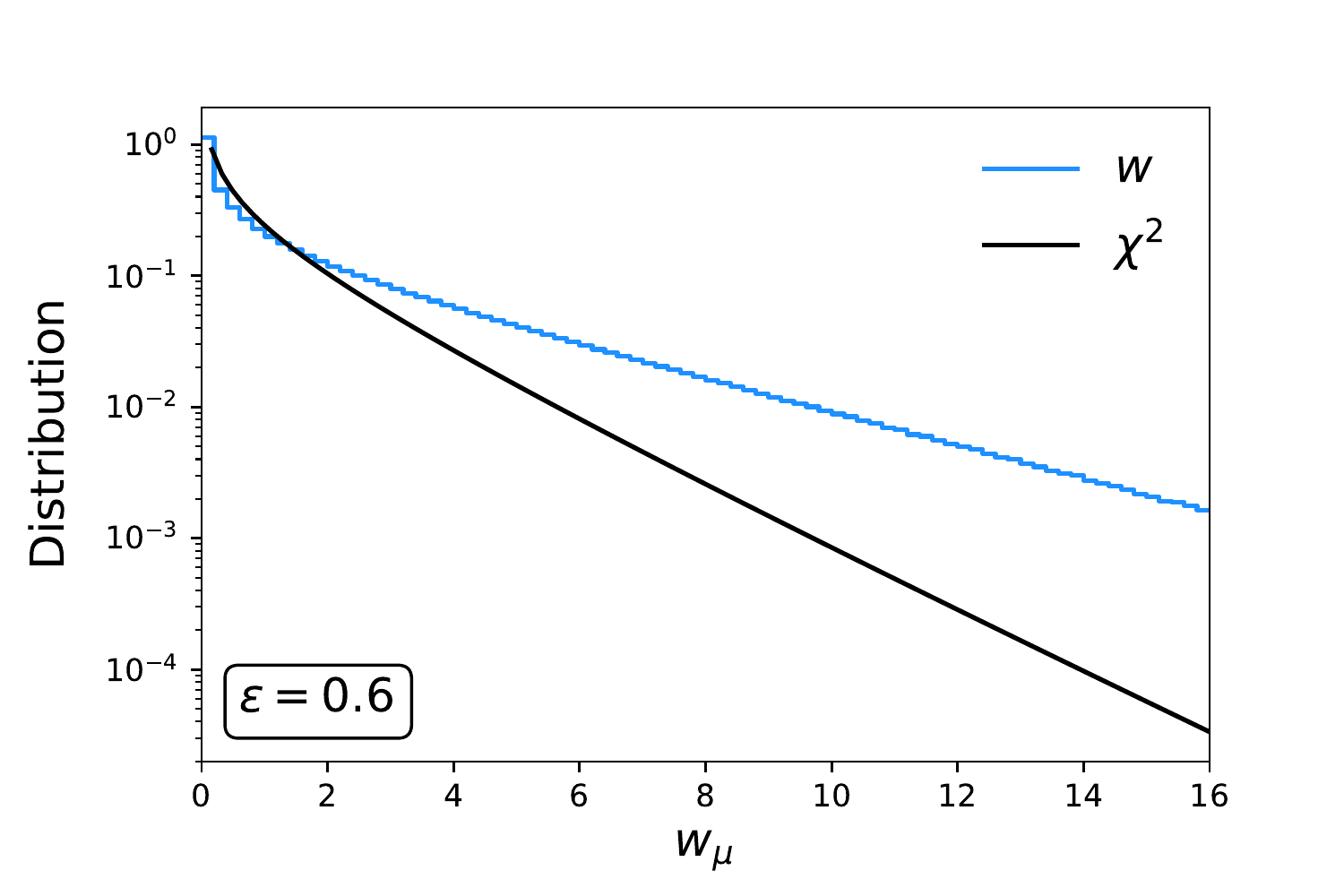}
    \caption{Distributions of $w_{\mu}$ (blue) for different values of the parameter $\varepsilon$ compared with the asymptotic $\chi^2$ distribution (black).}
    \label{fig:singMeas_LR_dist}
    \end{center}
\end{figure*}

Figure~\ref{fig:singMeas_LR_dist} shows the distributions of the likelihood ratio statistic with data generated according to Eq.~\eqref{eq:single_mes_model} setting $\mu=0$, $\sigma=1$ and $\varepsilon=0.01,\,0.2,\,0.4,\,0.6$. As found in Ref.~\cite{bib:Cowan2019}, the distribution deviates from the asymptotic $\chi_1^2$ form as the $\varepsilon$ parameter increases. The simple dependence of the single measurement model on the parameter $\varepsilon$ makes it an ideal candidate for studying the effectiveness of higher-order asymptotic methods in improving asymptotic formulae.

\subsection{Higher-order asymptotics for the single-measurement model}
\label{sec:hoasmm}

As one can see in Fig.~\ref{fig:singMeas_LR_dist}, the likelihood ratio exhibits noticeable deviations from its asymptotic $\chi_1^2$ distribution even for moderate values of $\varepsilon$.  It is therefore important to investigate whether higher-order statistics, namely $r^\ast_\mu$ and $w^\ast_\mu$ as defined in Eqs.~\eqref{eq:Rstar_def} and \eqref{eq:wstarDef}, can be better approximated by their asymptotic distributions, particularly for larger values of $\varepsilon$.


The asymptotic distribution of $r^\ast$ is a standard normal and it has an associated error term of $\mathcal{O}(n^{-3/2})$. For the single-measurement model, $\varepsilon$ gives the effective sample size ($n=1+1/2\varepsilon^2$), and thus the error term is expected to be of order $\mathcal{O}(\varepsilon^3)$ or smaller. In order to compute $r_\mu^\ast$ one needs $q_\mu$ as defined in Eq.~\eqref{eq:q1_def}.   The dependence of the likelihood of the single-measurement model on the data can be explicitly re-expressed in terms of the MLEs defined in Eq.\eqref{eq:inMes_MLE}:
\begin{equation}
\begin{split}
    \ell(\hat{\mu}, \widehat{\sigma^2}|\mu,\sigma^2)&=-\frac{1}{2}\frac{(\hat{\mu}-\mu)^2}{\sigma^2}- \left( \frac{1}{2}+\frac{1}{4\varepsilon^2} \right) \log{\sigma^2} \\ 
    &\phantom{{=}}-\frac{\widehat{\sigma^2}(1+2\varepsilon^2)}{4\varepsilon^2\sigma^2}\,.
\end{split}
\end{equation}
Therefore, it is possible to use Eq.\eqref{eq:q1_def} to compute  $q_\mu$, for which one finds
\begin{equation}
    q_\mu = \frac{\sqrt{(1+2\varepsilon^2) v}}{v+2\varepsilon^2(y-\mu)^2}(y-\mu)\,.
\end{equation}
Since the asymptotic distribution of $r_\mu^\ast$ is a standard normal, the asymptotic distribution of $r_\mu^{\ast2}$ is a $\chi_1^2$ distribution, and therefore it can be seen as a higher-order correction to the likelihood ratio. 

The second higher-order statistic we want to study is the Bartlett-corrected likelihood ratio,
\begin{equation}
     w_\mu^\ast = w_\mu\, \frac{M}{\text{E}[w_{\mu}]} \equiv  \frac{w_\mu}{1+b/M}\,,
\end{equation}
where E$[w_\mu] = M+b$ is what one must find to obtain the Bartlett correction. The Bartlett-corrected likelihood ratio $w_\mu^\ast$ is expected to be $\chi_1^2$ distributed in the asymptotic limit. The expectation value $\text{E}[w_\mu]$ can be estimated using the Lawley formula \eqref{eq:lawelyComposite_correction_def}, which yields
\begin{equation}
\label{eq:lawley_singMeasModel}
    \text{E}[w_\mu] = 1+ 3\varepsilon^2+\mathcal{O}(\varepsilon^4)\,.
\end{equation}
The asymptotic distribution of $w_\mu^\ast$ will have an error term of $\mathcal{O}(n^{-2})$, or equivalently $\mathcal{O}(\varepsilon^4)$ for the single-measurement model. All of the higher-order statistics described above, namely $r_\mu^{\ast2}$ and $w_\mu^\ast$, follow a $\chi^2$ distribution in the asymptotic limit. The expectation value in Eq.~\eqref{eq:lawley_singMeasModel} matches, at $\mathcal{O}(\varepsilon^2)$, the result found in \cite{bib:Cowan2019} using a Taylor expansion of the integral for its computation.

\begin{figure*}[h]
    \begin{center}
    \includegraphics[width=0.49\textwidth]{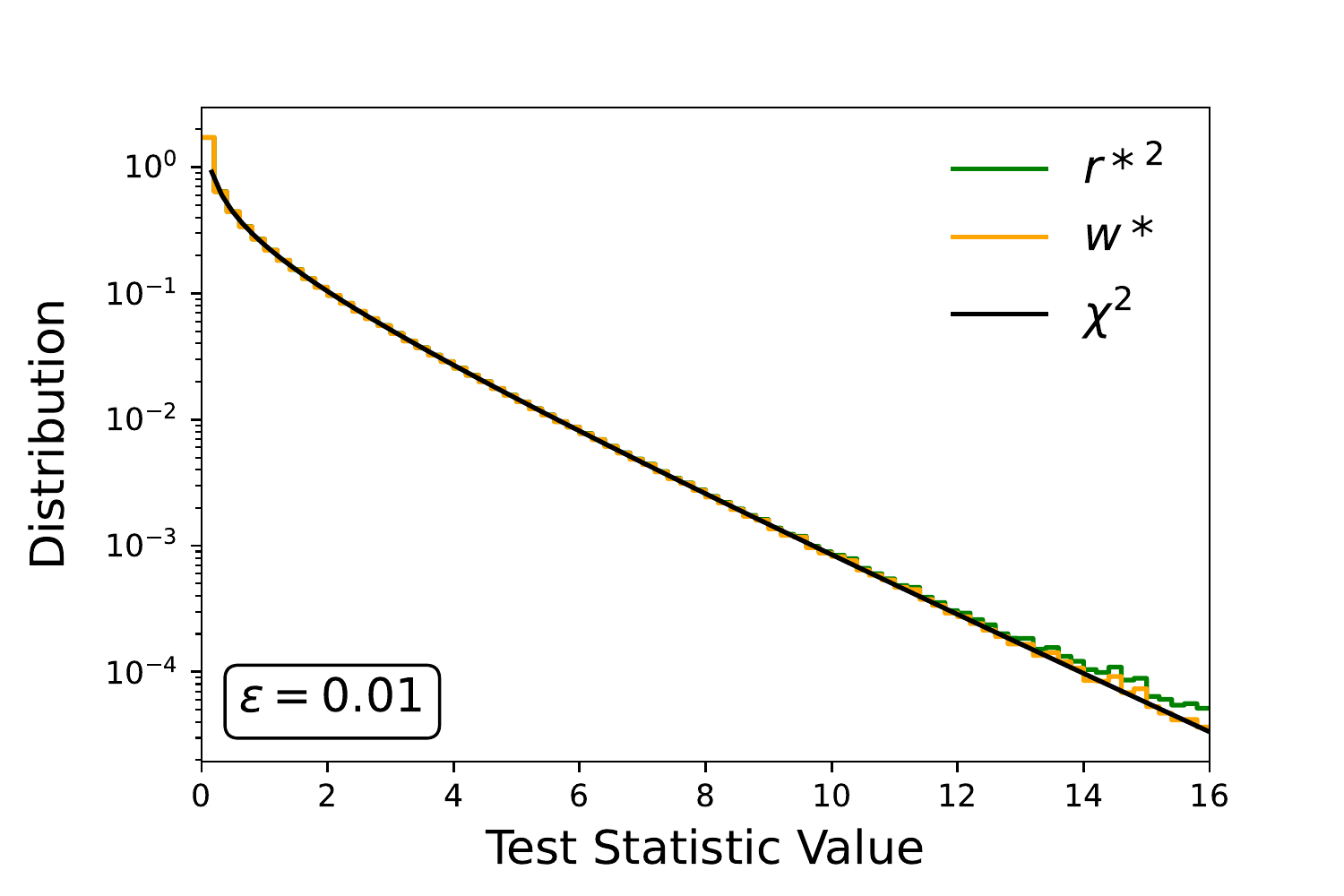}
    \includegraphics[width=0.49\textwidth]{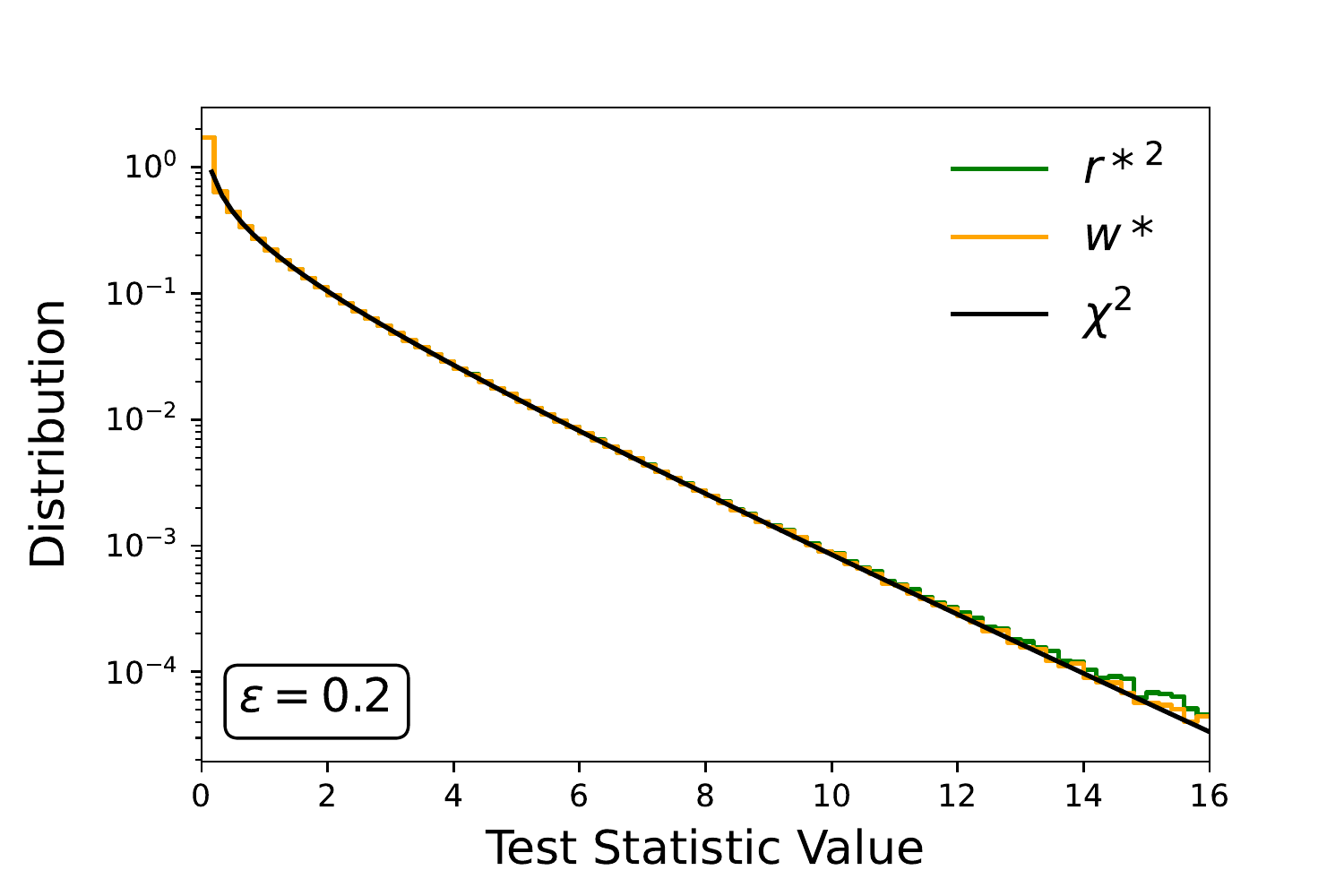}
    \includegraphics[width=0.49\textwidth]{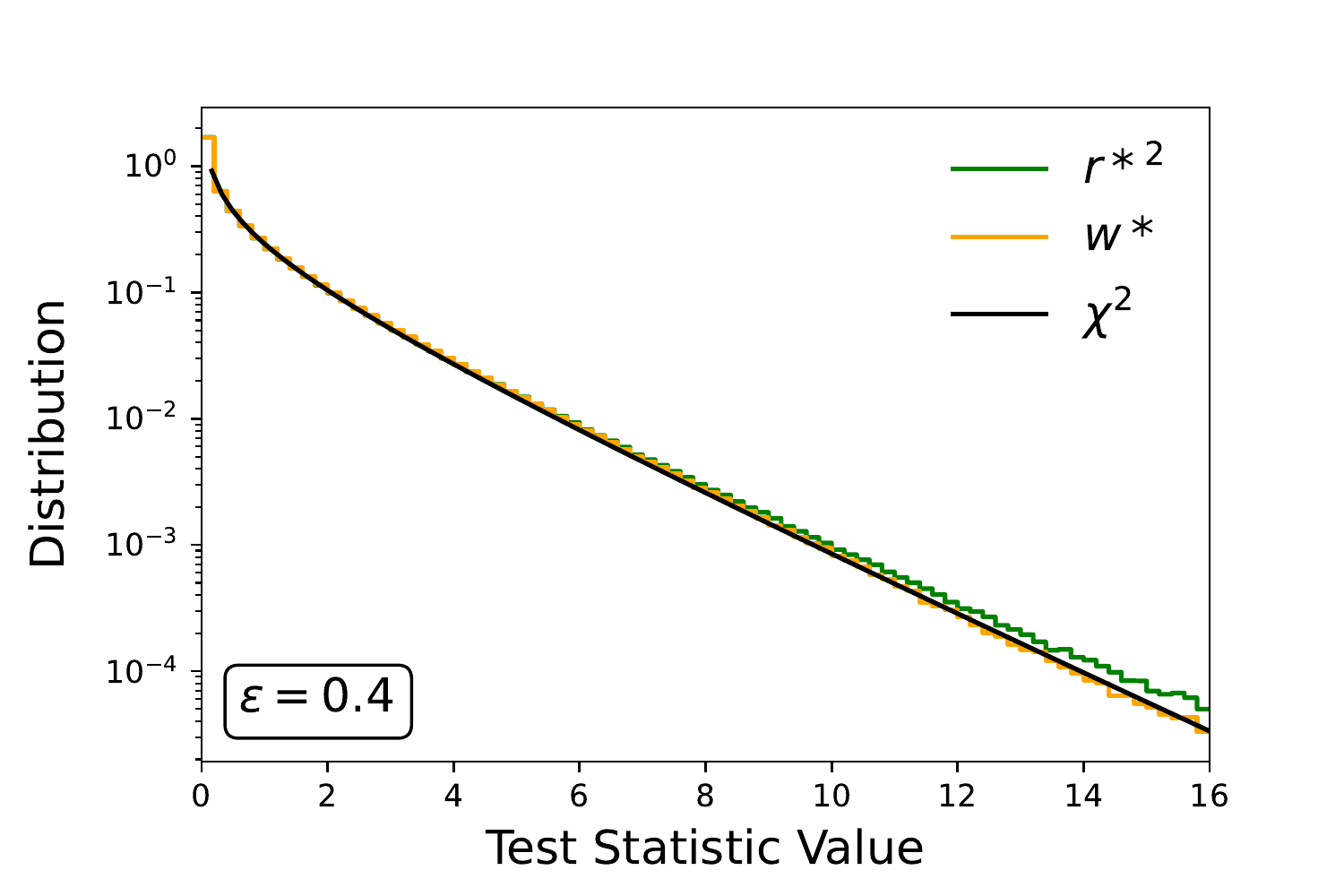}
    \includegraphics[width=0.49\textwidth]{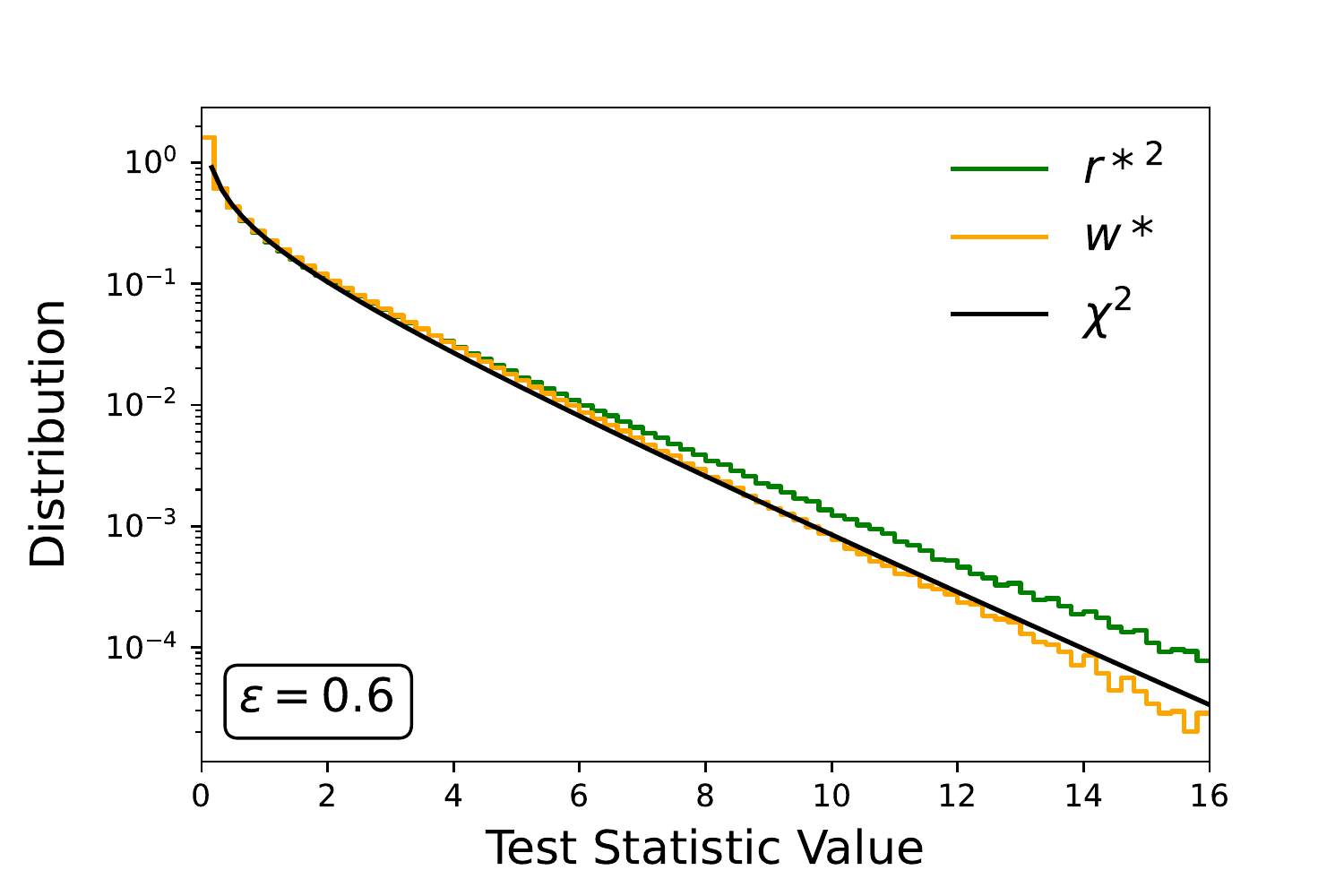}
    \caption{Distributions of $(r_\mu^{\ast})^2$ (green) and $w_\mu^\ast$ computed with the Lawley formula (orange) for different values of the parameter $\varepsilon$ compared with the $\chi^2$ asymptotic distribution (black).}
    \label{fig:singMeas_HOA_dist}
    \end{center}
\end{figure*}

In Fig.~\ref{fig:singMeas_HOA_dist}, we show the distributions of these two statistics for data generated according to Eq.~\eqref{eq:single_mes_model} with $\mu=0$, $\sigma=1$, and $\varepsilon$ values of 0.01, 0.2, 0.4, and 0.6. The distributions of two statistics are much better approximated by $\chi^2_1$ compared to the original likelihood ratio $w_\mu$, indicating that higher-order statistics provide significant improvements in this example.\\

\subsection{Confidence intervals for the single-measurement model}
\label{sec:cismm}
The likelihood ratio is a commonly used tool for deriving confidence regions, typically obtained by finding the $p$-value of $\boldsymbol{\mu}$ and then solving the equation $p_{\boldsymbol{\mu}}=\alpha$, where $1-\alpha$ represents the desired confidence level. In the case of the single-measurement model, which involves only one parameter of interest $\mu$, our goal is to construct a confidence interval for it as described in Sec.~\ref{sec:first_order_theory}. To obtain the $p$-value, the distribution of $w_\mu$ must be determined.  As seen in Fig.~\ref{fig:singMeas_LR_dist}, the distribution of $w_{\mu}$ departs from its asymptotic form for large values of $\varepsilon$, and $p$-values derived from a $\chi^2_1$ distribution are thus not accurate.
To address this, we can use higher-order statistics such as $r_\mu^\ast$ and $w_\mu^\ast$ to compute the $p$-values, i.e.,
\begin{equation}
    p_{\mu} = \int_{w^\ast_{ \text{obs}}}^{\infty}f_{\chi_1^2}(w^\ast) \, dw^\ast= 1 - F_{\chi_1^2}[w^\ast_{ \text{obs}}]\,,
\end{equation}
or
\begin{equation}
    p_{\mu} = \int_{r^{\ast2}_{ \text{obs}}}^{\infty}f_{\chi_1^2}(r^{\ast2}) \, dr^{\ast2}= 1 - F_{\chi_1^2}[r^{\ast2}_{ \text{obs}}]\,.
\end{equation}

\begin{figure*}[t]
    \begin{center}
    \includegraphics[width=0.6\textwidth]{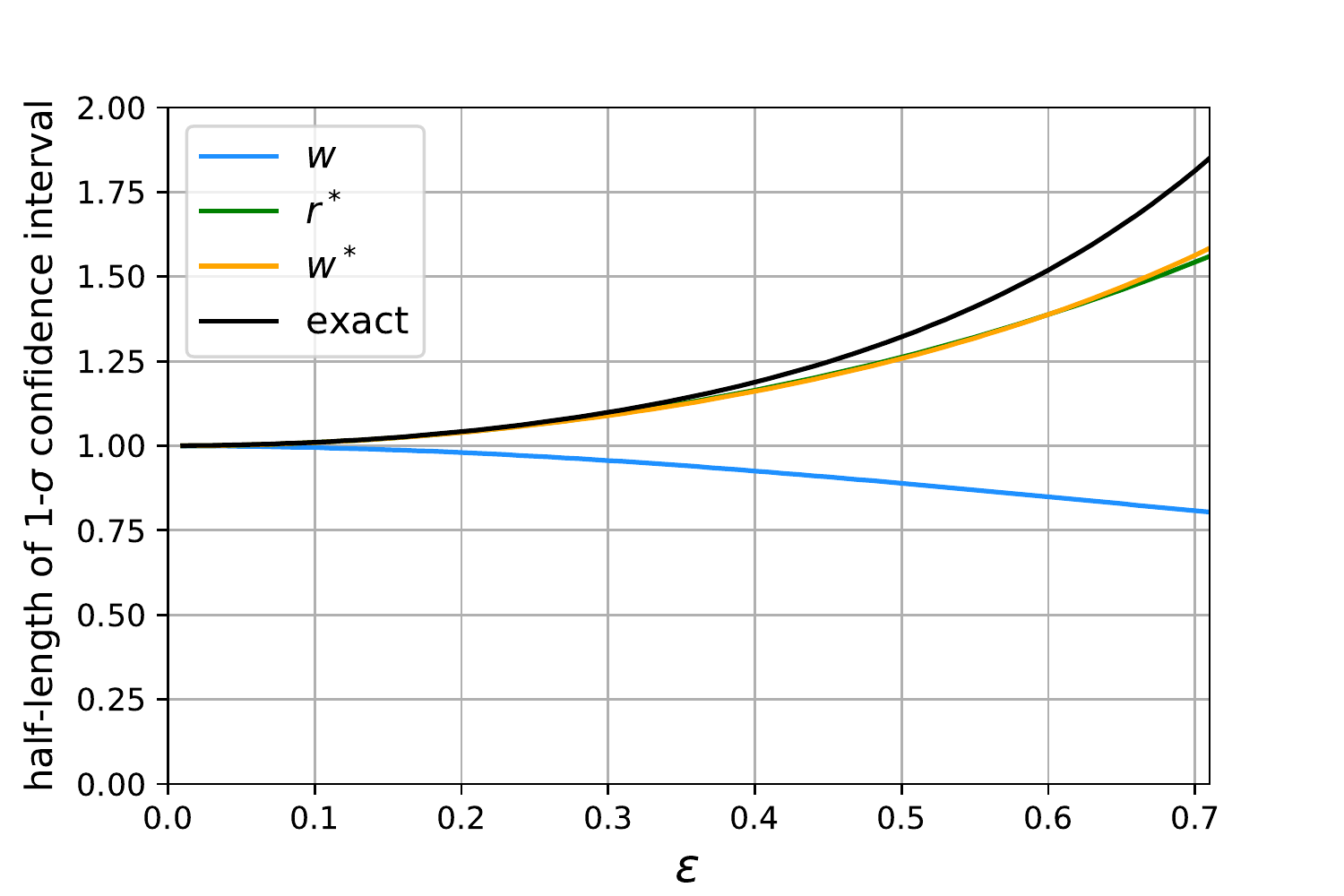}
    \caption{Half-length of $1$-$\sigma$ confidence intervals ($68.3\%$ confidence level) for $\mu$ (Eq.~\eqref{eq:single_mes_model}) as a function of $\varepsilon$, computed using $w_\mu$ (blue), $r^\ast_\mu$ (green) and $w^\ast_\mu$ calculated using the Lawley formula (orange). The black curve represents the exact half-length of the confidence interval.}
    \label{fig:singMeas_CI}
    \end{center}
\end{figure*}

To illustrate this we find the confidence interval for $\mu$ as a function of the parameter $\varepsilon$ under the assumption that the observed values of $y$ and $v$ are $0$ and $1$, respectively. Figure~\ref{fig:singMeas_CI} shows a comparison of the confidence intervals obtained using the likelihood ratio $w_\mu$ and the higher-order statistics $r_\mu^\ast$ and $w_\mu^\ast$. Additionally, the confidence interval is also computed by calculating the $p$-value exactly, as described in \cite{bib:Cowan2019}. The plot in Fig.~\ref{fig:singMeas_CI} shows that the use of higher-order statistics significantly improves the accuracy of the confidence interval. Figure \ref{fig:singMeas_CI} confirms the findings in \cite{bib:Cowan2019}, but also shows that $r^\ast$ provides estimates of the confidence interval as accurate as those computed with $w^\ast$, proving to be a valid alternative to the Bartlett correction.

\section{Simple-average model}
\label{sec:sam}

The single-measurement model can be extended to an average of $N$ measurements $\boldsymbol{y}=(y_1,\,...,\,y_N)$, which are assumed to follow a Gaussian distribution with a common mean $\mu$ and variances $\boldsymbol{\sigma^2}=(\sigma_1^2,\,...,\,\sigma_N^2)$.  Here the variances are assumed to be uncertain with independent estimates $\boldsymbol{v}=(v_1,\,...,\,v_N)$ that are gamma distributed with error-on-error parameters (cf.\ Eq.~\eqref{eq:eps_def}) of  $\boldsymbol{\varepsilon}=(\varepsilon_1,\,...,\,\varepsilon_N)$. The likelihood of the model is thus
\begin{equation}
\begin{split}
    L(\mu, \boldsymbol{\sigma^2}) =&\prod_i^{N}\frac{1}{\sqrt{2\pi\sigma_i^2}}e^{-(y_i-\mu)^2/2\sigma_i^2}\\
    & \times \prod_i^{N}\frac{\beta_i}{\Gamma(\alpha_i)}v^{\alpha_i-1}e^{-\beta_i v_i}\,,
\end{split}
\end{equation}
where $\alpha_i=1/4\varepsilon_i^2$ and $\beta_i=1/(4 \varepsilon_i^2\sigma_i^2)$. Equivalently, the log-likelihood of the model is 
\begin{equation}
\label{eq:simpleAverage_logLik}
\small
    \ell(\mu, \boldsymbol{\sigma^2})=-\frac{1}{2}\sum_{i=1}^N \left[\frac{(y_i-\mu)^2}{\sigma_i^2}+\left(1+\frac{1}{2\varepsilon_i^2}\right)\log{\sigma_i^2}+\frac{v_i}{2\varepsilon_i^2\sigma_i^2} \right] \,.
\end{equation}
In contrast to the full Gamma Variance Model described in Sec.~\ref{sec:gammaModel_def}, it does not include nuisance parameters 
$\theta_i$ or their estimates, but rather treats the variances $\sigma_i^2$ of the primary measurements $y_i$ as uncertain.
It can be easily generalized to a curve-fitting problem where the expectation value of each measurement $y_i$ can be defined as a function of the parameters of interest $\boldsymbol{\mu}$ and a control measurement $x_i$, i.e., $\text{E}[y_i]=f(x_i; \boldsymbol{\mu})$. 

The log-likelihood of Eq.~\eqref{eq:simpleAverage_logLik} profiled over the $\boldsymbol{\sigma^2}$ is given by
\begin{equation}
\label{eq:simpleAverage_profLogLik}
    \ell_p(\mu)=-\frac{1}{2}\sum_i^N\left(1+\frac{1}{2\varepsilon_i^2}\right)\log\left[{1+2\varepsilon_i^2\frac{\left(y_i-\mu\right)^2}{v_i}}\right]\,,
\end{equation}
which has been computed using the profiled value of $\sigma_i^2$:
\begin{equation}
    \widehat{\widehat{\sigma_i^2}}=\frac{v_i+2\varepsilon_i^2(y_i-\mu)^2}{1+2\varepsilon_i^2}\,.
\end{equation}

As in the example of the single-measurement model from Sec.~\ref{sec:smm}, we compute the likelihood ratio $w_\mu$, and the statistics $r^\ast_\mu$ and $w^\ast_\mu$.  These require the MLE $\hat{\mu}$, which in general must be found numerically.
%
As discussed in Sec.~\ref{sec:gammaModel_def}, the distribution of the likelihood ratio $w_\mu$ is expected to deviate from its asymptotic $\chi^2_1$ form, and we investigate whether the higher-order statistics $r_\mu^\ast$ and $w_\mu^\ast$ can improve the precision of the inference on $\mu$.

Because the log-likelihood of Eq.\eqref{eq:simpleAverage_logLik} cannot be written explicitly as a function of the MLEs, but is known in terms of the data values $y_i$, the correction term $q_{\mu}$, needed to compute $r_\mu^\ast$,
must be found using Eq.\eqref{eq:q2_def}. Additionally, to compute $q_\mu$, as discussed in Sec.~\ref{sec:p_star_aprox}, a vector of pivotal quantities $\boldsymbol{z}=(z_{y_1},\,...,\,z_{y_N},\,z_{v_1},\,...,\,z_{v_N})$ must be defined and used in Eqs.~\eqref{eq:canonical_params_def} and~\eqref{eq:Vmatrix_def}.  We choose these to be 
\begin{equation}
\begin{split}
    &z_{y_{i}} = \frac{(y_{i}-\mu)^2}{\sigma_{i}^2}\sim \chi^2_1\,,\\ 
    &z_{v_{i}} = \frac{v_i}{\sigma_{i}^2}\sim\chi^2_1\,.
\end{split}
\end{equation}

The Bartlett-corrected likelihood ratio $w_\mu^\ast$ can be calculated using the Lawley formula \eqref{eq:lawelyComposite_correction_def}. The result can be expanded at order $\varepsilon_i^2$ as
\begin{equation}
\label{eq:lawley_sam}
\begin{split}
    \text{E}[w_\mu]&= 1+ \frac{4}{\sum_{i=1}^N 1/v_i}\sum_{i=1}^N\frac{\varepsilon_i^2}{v_i}\\ & \phantom{{=}}- \frac{1}{(\sum_{i=1}^N 1/v_i)^2}\sum_{i=1}^N\frac{\varepsilon_i^2}{v_i^2}+\sum_{i=1}^N\mathcal{O}(\varepsilon_i^4)\,,
\end{split}
\end{equation}
which in the limit $\varepsilon_i \rightarrow 0$ gives $1$ as expected.  Using this result one can thus find the corrected statistic $w_{\mu}^{\ast} = w_{\mu} / \text{E}[w_{\mu}]$.

\subsection{Confidence intervals for the parameter of interest}
\label{sec:cipoi}



In this section, we compute the $68.3\%$ confidence interval for the parameter $\mu$ of Eq.~\eqref{eq:simpleAverage_logLik} to benchmark the higher-order statistics computed in the last section. Specifically, we consider the simple case of averaging two measurements, $y_1$ and $y_2$, with observed values of $+\delta$ and $-\delta$, respectively.  
The estimates of the variances $v_1$ and $v_2$ are set to $1$.  We consider $\delta = 0.5$, which represents the case where $y_1$ and $y_2$ are reasonably consistent, and $\delta = 1.5$, corresponding to a substantial tension between the two measurements. Both measurements are assigned equal error-on-error parameters, $\varepsilon_1=\varepsilon_2=\varepsilon$, and we present results as a function of $\varepsilon$.  The intervals are found using the likelihood ratio $w_\mu$ and the higher-order statistics $r_\mu^\ast$ and $w_\mu^\ast$. 

In addition, the confidence interval is found by estimating the $p$-value of the likelihood ratio using MC. This is done by generating the exact distribution of the data for a fixed value of $\mu$ while setting the nuisance parameters $\sigma_i^2$ to their profiled values. In Particle Physics, this technique is commonly known as the \textit{profile construction} \cite{bib:Cranmer2005} or \textit{hybrid resampling} \cite{bib:Chuang2000, bib:BodhisattvaSen2009} method (in statistics often referred to as a \textit{parametric bootstrap}). This technique is expected to provide the most accurate determination of the intervals that is computationally feasible.

\begin{figure*}[t]
    \begin{center}
    \includegraphics[width=0.6\textwidth]{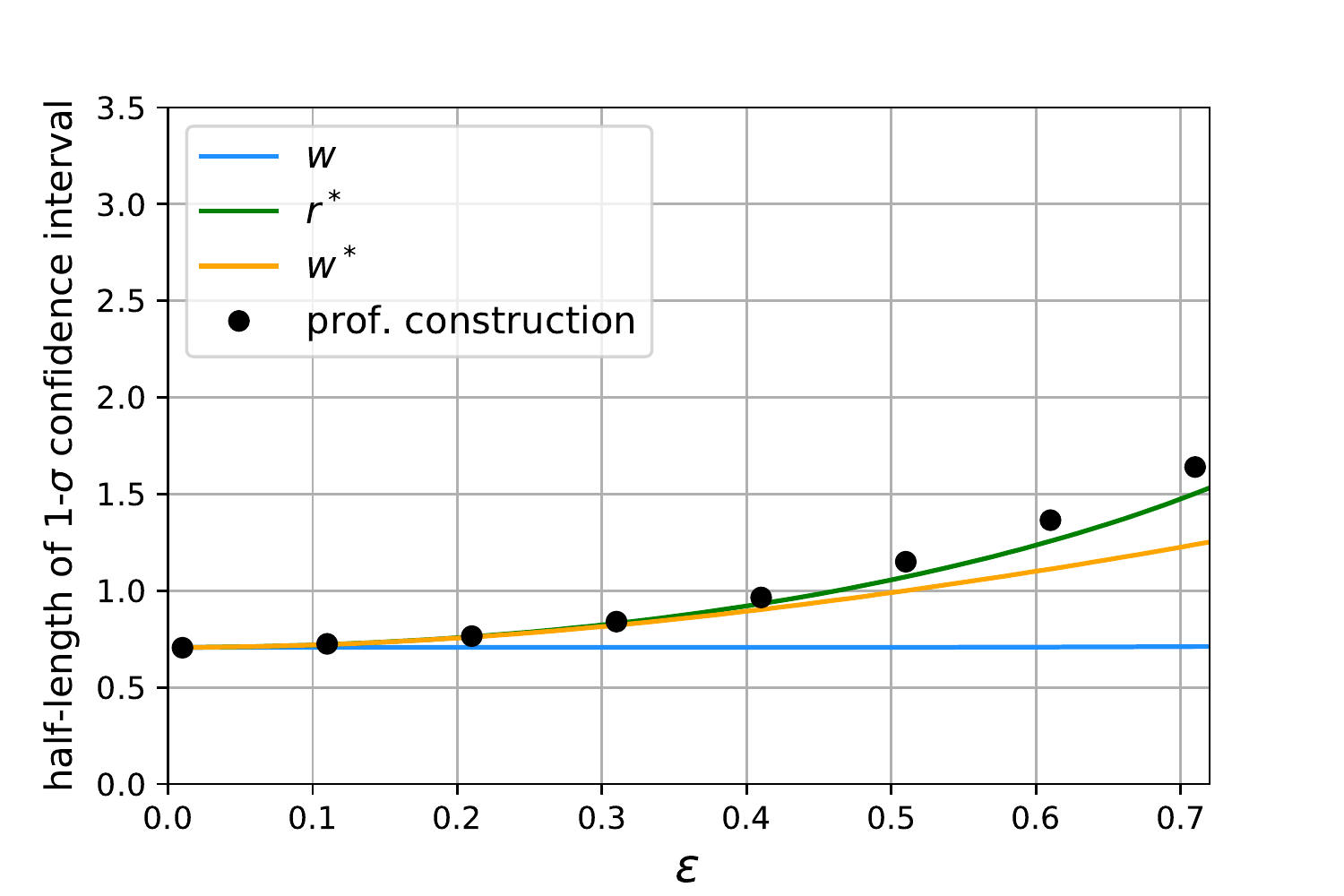}
    \includegraphics[width=0.6\textwidth]{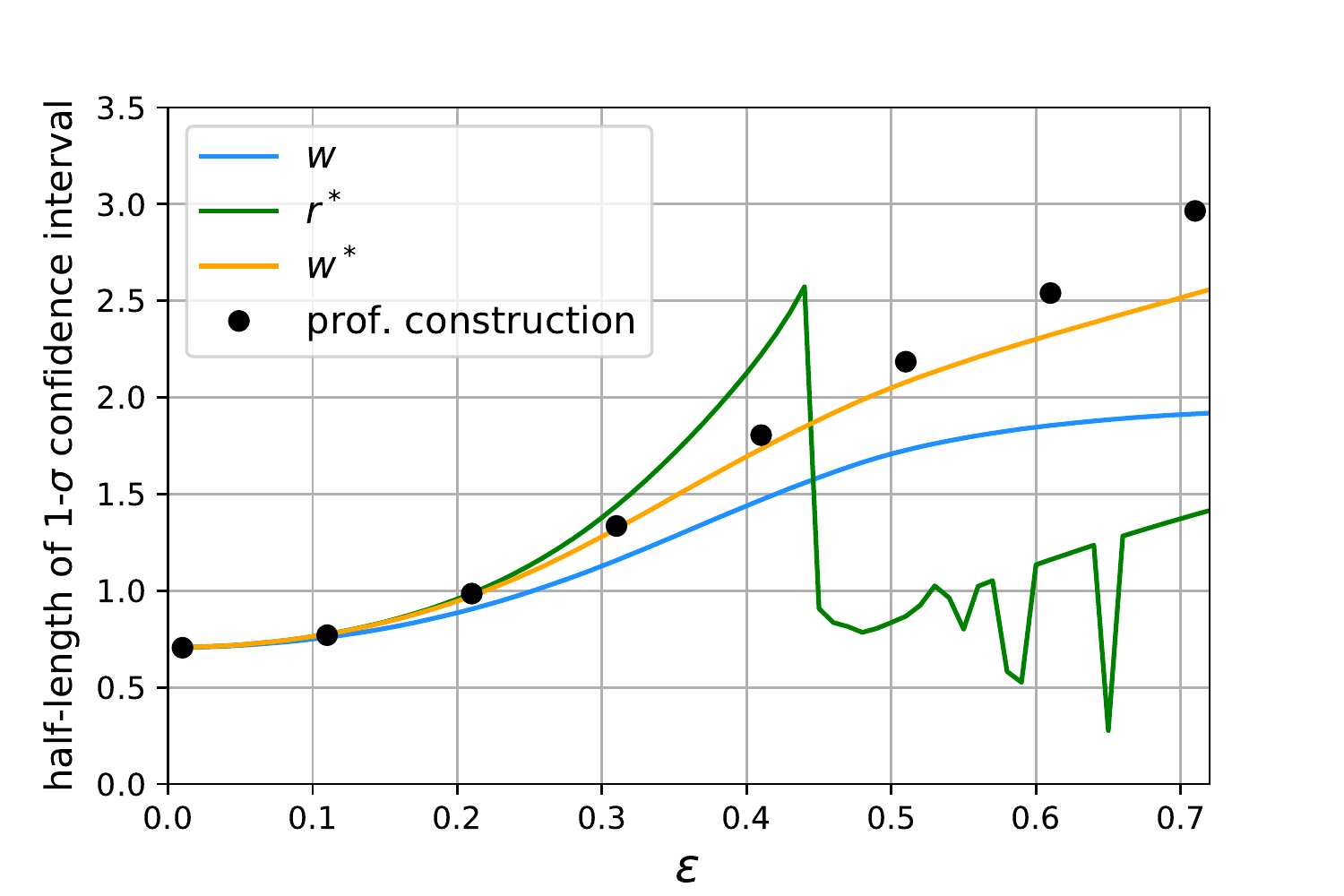}
    \caption{Half-length of $1$-$\sigma$ confidence intervals for parameter $\mu$ (Eq.~\eqref{eq:simpleAverage_logLik}) as a function of $\varepsilon$ for $\delta=0.5$ (top) and $\delta=1.5$ (bottom), computed using $w_\mu$ (blue), $r^\ast_\mu$ (green) and $w^\ast_\mu$ calculated using the Lawley formula (orange). The black dots represent our most precise estimate of the interval, computed using profile construction. The confidence interval obtained with $r^\ast_\mu$ becomes numerically unstable for large values of $\varepsilon$ when there is significant tension in the dataset (bottom).}
    \label{fig:simpleAverage_CI}
    \end{center}
\end{figure*}

 The uncorrected interval based on the likelihood ratio $w_{\mu}$ is found to undershoot the result from profile construction in both cases, with the discrepancy increasing with $\varepsilon$. Conversely, the intervals obtained using the $r_\mu^\ast$ and $w^\ast$ statistics are found to be in good agreement with that from profile construction when the averaged data are mutually compatible (top panel of Fig.\eqref{fig:simpleAverage_CI}). However, for larger values of $\varepsilon$, $r_\mu^\ast$ breaks down as the tension in the observed data grows (see the lower panel of Fig.\eqref{fig:simpleAverage_CI}), leading to numerical instability. This occurs because the correction term $q_\mu$ in the definition of $r_\mu^\ast$ (see Eq.~\eqref{eq:Rstar_def}) no longer effectively corrects the likelihood root, as the likelihood becomes strongly non-Gaussian. Consequently, deviations from the asymptotic limit can no longer be adequately addressed by a correction term.

A conservative approach to determine the applicability of $r_\mu^\ast$ in improving the likelihood ratio predictions is to verify whether the arguments of the logarithmic terms of Eq.~\eqref{eq:simpleAverage_profLogLik} are within their radius of convergence. Specifically, for the endpoints of the confidence interval, one should check whether the inequality
\begin{equation}
2\varepsilon_i^2\frac{\left(y_i-\mu\right)^2}{v_i}<1
\end{equation}
is satisfied for every measurement $y_i$. In the example above, this condition implies that one should not trust the accuracy of the result from $r_\mu^\ast$ if $\varepsilon\geq0.5$ for $\delta=0.5$ and $\varepsilon\geq0.3$ for $\delta=1.5$.

\subsection{Goodness-of-fit}
\label{sec:simpleAverage_gof}

A confidence region for the parameters of interest does not, on its own, provide a measure of how well the selected model describes the observed data.  This can be quantified using a goodness-of-fit statistic for which we take 
\begin{equation}
    \label{eq:simpleAverage_GoF}
    q = -2\log\frac{L(\hat{\mu},\hat{\boldsymbol{\sigma}}^2)}{L_{\rm s}(\hat{\boldsymbol{\varphi}},\hat{\boldsymbol{\sigma}}^2)}\,,
\end{equation}
where $L_{\rm s}$ represents the likelihood of the \textit{saturated model}. This is obtained by replacing the expectation values $\text{E}[y_i]=\mu$ with a set of independent parameters $\boldsymbol{\varphi}=(\varphi_1,\,,...,\,,\varphi_N)$, such that $\text{E}[y_i]=\varphi_i$. Since there is an adjustable $\varphi_i$ for each measurement $y_i$, one finds $\hat{\varphi}_i = y_i$, $\hat{\sigma}^2_i = v_i$, and the goodness-of-fit statistic reduces to
\begin{equation}
    q = 
    \sum_{i=1}^N\left(1+\frac{1}{2\varepsilon_i^2}\right)\log\left[{1+2\varepsilon_i^2\frac{\left(y_i-\hat{\mu}\right)^2}{v_i}}\right]\,.
\end{equation}

If the above expression is expanded in powers of $\varepsilon_i^2$, in the limit $\varepsilon_i^2 \rightarrow 0$ one finds
\begin{equation}
    q  =\sum_{i=1}^N\frac{\left(y_i-\hat{\mu}\right)^2}{v_i}+\mathcal{O}_p(\varepsilon_i^2)\,.
\end{equation}
In this limit, $q$ reduces to a sum of squares of Gaussian-distributed quantities, and thus its distribution follows a chi-square distribution with $N-1$ degrees of freedom. 
However, for large values of the $\varepsilon_i$ parameters, deviations from the $\chi^2_{N-1}$ asymptotic distribution are expected.  

To correct the goodness-of-fit statistic using higher-order asymptotics, $q$ needs to be defined as a likelihood ratio. This can be done by defining the saturated model such that the simple-average model is nested within it. A possible choice is to define the saturated model as
\begin{equation}
\begin{split}
    &\ell_{\rm s}(\boldsymbol{\alpha}, \mu, \boldsymbol{\sigma^2})=\log L_{\rm s}(\boldsymbol{\alpha}, \mu, \boldsymbol{\sigma^2}) \\*[0.2cm]
    & =-\frac{1}{2}\left[\sum_i^N\frac{(y_i-\alpha_i-\mu)^2}{\sigma_i^2}+\left(1+\frac{1}{2\varepsilon_i^2}\right)\log{\sigma_i^2}+\frac{v_i}{2\varepsilon_i^2\sigma_i^2}\right]\,,
\end{split}
\end{equation}
where we fix $\alpha_N=-\sum_{i=1}^{N-1}\alpha_i$ so that $\sum_{i=1}^{N}\alpha_i=0$. Given this definition, the simple-average model is recovered by fixing all the $\alpha_i$ to zero, hence $\ell=\ell_{\rm s}(\boldsymbol{\alpha}=\boldsymbol{0}, \mu, \boldsymbol{\sigma^2})$. Therefore, the goodness-of-fit statistic can be written as a likelihood ratio of the saturated model,
\begin{equation}
    q = -2\log\frac{L_{\rm s}(\boldsymbol{\alpha}=\boldsymbol{0},\hat{\hat{\mu}},\hat{\hat{\boldsymbol{\sigma}}}^2)}{L_{\rm s}(\hat{\boldsymbol{\alpha}},\hat{\mu},\hat{\boldsymbol{\sigma}}^2)}\,,
\end{equation}
and its Bartlett correction can be computed using the Lawley formula. This is done by treating the $\alpha_i$ as parameters of interest and $\mu$ and the $\sigma_i^2$ as nuisance parameters. The result is given by:
\begin{equation}
\label{eq:lawley_q}
\begin{split}
    &\text{E}[q] =  \,N - 1\\ 
    &- 8\sum_{i=1}^N\sum_{j=1}^{N-1}k^{\mu\alpha_j}\,C_{ij}\frac{r^2_i}{v_i} - 4\sum_{i=1}^N\sum_{j,k=1}^{N-1}k^{\alpha_j\alpha_k}\,C_{ij}C_{ik} \frac{r^2_i}{v_i}  \\
    &-4\sum_{i=1}^N\sum_{j,k=1}^{N-1}k^{\mu\alpha_j}k^{\mu\alpha_k}\,C_{ij}C_{ik}\frac{r^2_i}{v^2_i} \\
    &-2\sum_{i=1}^N\sum_{j,k,p=1}^{N-1}k^{\mu\alpha_j}k^{\alpha_k\alpha_p}\,C_{ij}C_{ik}C_{ip}\frac{r^2_i}{v^2_i}\\
    &-\sum_{i=1}^N\sum_{j,k,p,q=1}^{N-1}k^{\alpha_j\alpha_k}k^{\alpha_p\alpha_q}\,C_{ij}C_{ik}C_{ip}C_{iq}\frac{r^2_i}{v^2_i}\,.
\end{split}
\end{equation}
The terms $k^{\alpha_i\alpha_j}$ and $k^{\mu\alpha_i}$ refer to the components of the inverse of the expectation value of the Hessian matrix of the likelihood (see Eq.~\eqref{eq:k-terms1_def}) and are computed for $\sigma^2_{y_i}=v_i$. The matrix $C$ is a $N\times (N-1)$ matrix, whose only non-zero entries are:
\begin{equation}
\begin{split}
    &C_{ij} = 1  \quad \forall \,i = j\,,\\
    &C_{Nj} = -1  \quad \forall \,j \,.
\end{split}
\end{equation}
The $r^\ast$ statistic cannot be applied to goodness-of-fit unless $N=2$, as it can only be computed for models with one parameter of interest.

To measure how well the model describes the observed data, one can compute the $p$-value of the goodness-of-fit,
\begin{equation}
    p = \int_{q_{ \text{obs}}}^{\infty} f(q) \, dq= 1 - F[q_{\rm obs}]\,,
\end{equation}
In general, small values of the $p$-value are associated with a bad agreement between the model and the data. In Particle Physics, $p$-values are typically converted to a related quantity $Z$ called the \textit{significance}, defined (for a two-sided test) as 
\begin{equation}
\label{eq:significance}
    Z=\Phi^{-1}(1-p/2)\,,
\end{equation}
where $\Phi^{-1}$ is the inverse cumulative distribution of a standard normal. The $p$-value is thus equated to the probability of a Gaussian variable to fluctuate $Z$ standard deviations or more away from the mean (i.e., a probability of $p/2$ in each direction).

\begin{figure*}[t]
    \begin{center}
    \includegraphics[width=0.6\textwidth]{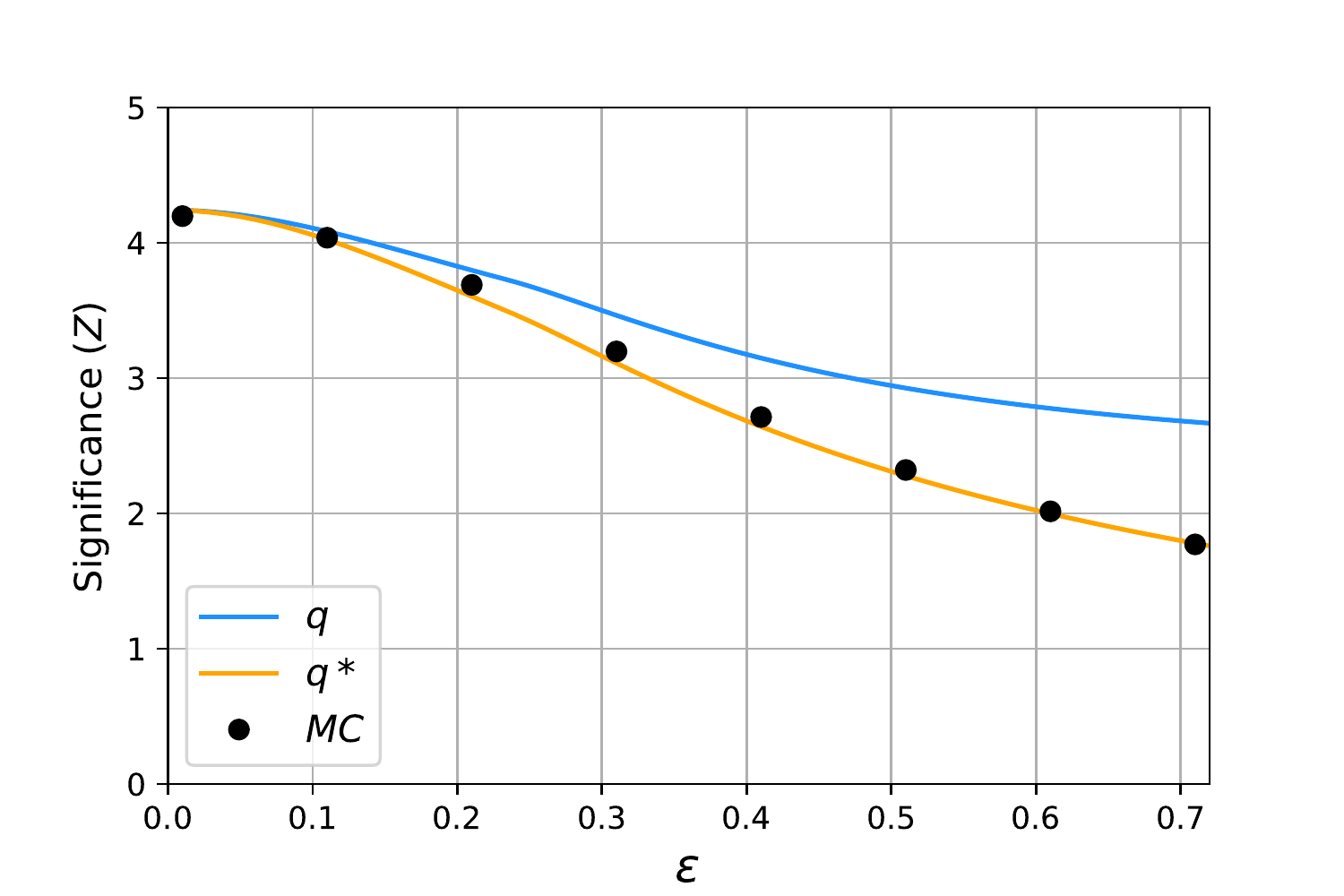}
    \caption{Significance of discrepancy for $\delta=3$ as defined by Eq.~\eqref{eq:significance} plotted as a function of $\varepsilon$. Higher significance values indicate greater tension between the data. The significance was computed using the goodness-of-fit statistic (blue) defined by Eq.~\eqref{eq:simpleAverage_GoF} and its Bartlett corrected counterpart calculated using the Lawley formula (orange). The black dots represent the significance computed by generating the distribution of $q$ with MC. The plot illustrates that the Bartlett correction enables an accurate approximation of the numerical predictions.}
    \label{fig:simpleAverage_gof}
    \end{center}
\end{figure*}

To illustrate this method, we find the $p$-value and corresponding significance $Z$ for an average of two incompatible measurements, namely $y_1=-3$ and $y_2=3$, with estimated variances of $v_1=v_2=1$. We set $\varepsilon_1=\varepsilon_2=\varepsilon$ and we show the results as a function of $\varepsilon$. The significance is estimated using both the goodness-of-fit statistic $q$ and the Bartlett-corrected version $q^\ast$. Figure~\ref{fig:simpleAverage_gof} shows the significance as a function of the parameter $\varepsilon$. The use of the Bartlett correction results in a substantial improvement in estimating the significance, almost perfectly overlapping with the MC estimates. This is a crucial result, as estimating $p$-values of goodness-of-fit for incompatible data can require a large number of pseudo-experiments. For example, roughly on the order of $10^5$ simulated experiments are necessary to accurately capture a $4\sigma$ effect.

The Bartlett correction can also be found using MC to estimate $\text{E}[w_{\mu}]$, which takes less time than computing the full distribution. The result is very close to what is found from the Lawley formula, and the latter has the advantage of avoiding MC entirely.

\section{Averages using the full Gamma Variance Model}
\label{sec:GVM_ave}

The simple-average model of the previous section assumes that the individual measurements are unbiased estimators of the parameter of interest, i.e., $\text{E}[y_i] = \mu$, but that the standard deviations of the $y_i$ are uncertain.   In practice, the $\sigma_{y_i}$ are often well estimated because they correspond to the statistical uncertainty in $y_i$, and thus they are directly related to a sample size or a number of counts.  It often happens, however, that  $y_i$ may have a potential bias which must be constrained with a control measurement, as described in the full Gamma Variance Model of Sec.~\ref{sec:gammaModel_def}.  In this section we apply higher-order asymptotic corrections to this case.

More precisely, $N$ measurements are assumed to be independent and Gaussian distributed with means $\text{E}[y_i]=\mu+\theta_i$ and known variances (the ``statistical errors'') $\text{V}[y_i]=\sigma_{y_i}^2$. Here, the nuisance parameters $\theta_i$ represent potential biases to the means of the $y_i$.  As described in Sec.~\ref{sec:gammaModel_def}, their values are estimated with independent Gaussian distributed control measurements $u_i$, whose variances $\sigma_{u_i}^2$ (the ``systematic errors'') are treated as adjustable parameters.  The $\sigma_{u_i}^2$ are estimated by
measurements $v_i$, whose gamma distributions are
characterized by the error-on-error parameters $\varepsilon_i$.
The log-likelihood of the model is
\begin{equation}
\label{eq:GVMaverage_logLik}
\begin{aligned}
    \ell(\mu, \boldsymbol{\theta}, \boldsymbol{\sigma_{u}^2})=&-\frac{1}{2}\sum_{i=1}^N
    \left[ 
    \frac{(y_i-\mu-\theta_i)^2}{\sigma_{y_{i}}^2}+\frac{(u_i-\theta_i)^2}{\sigma_{u_{i}}^2}\right.\\
    &\left.+\left(1+\frac{1}{2\varepsilon_i^2}\right)\log{\sigma_{u_{i}}^2}+\frac{v_i}{2\varepsilon_i^2\sigma_{u_{i}}^2} \right] \,,
\end{aligned}
\end{equation}
and the profiled log-likelihood $\ell_p$ can be computed using
\begin{equation}
    \widehat{\widehat{\sigma_{u_i}^2}}=\frac{v_i+2\varepsilon_i^2(u_i-\theta_i)^2}{1+2\varepsilon_i^2}\,,
\end{equation}
leading to
\begin{equation}
\begin{aligned}
    \ell_p(\mu, \boldsymbol{\theta}) = &-\frac{1}{2}\sum_{i=1}^N \left[ \frac{(y_i-\mu-\theta_i)^2}{\sigma_{y_{i}}^2}\right.\\
    &\left.+\left(1+\frac{1}{2\varepsilon_i^2}\right)\log\left(1+2\varepsilon_i^2\frac{(u_i-\theta_i)^2}{v_i} \right) \right]\,.
\end{aligned}
\end{equation}
The MLEs $\hat{\mu}$ and $\hat{\theta_i}$ can be found numerically or by solving a system of cubic equations (see Ref.~\cite{bib:Cowan2019}).

As before, we compute the likelihood ratio $w_\mu$ and the higher-order statistics $r^\ast_\mu$ and $w^\ast_\mu$. To compute $r_\mu^\ast$, one needs to calculate $q_\mu$ as defined in Eq.~\eqref{eq:q2_def}. This requires a vector of pivotal quantities \\$\boldsymbol{z}=(z_{y_1},\,...,\,z_{y_N},\,z_{u_1},\,...,\,z_{u_N},\,z_{v_1},\,...,\,z_{v_N})$, which can be defined as 
\begin{equation}
\begin{split}
    &z_{y_{i}} = \frac{(y_{i}-\mu-\theta_i)^2}{\sigma_{y_i}^2}\sim \chi_1^2\,,\\ 
    &z_{u_{i}} = \frac{(u_{i}-\theta_i)^2}{\sigma_{u_i}^2}\sim \chi_1^2\,,\\
    &z_{v_{i}} = \frac{v_i}{\sigma_{u_i}^2}\sim \chi_1^2\,.
\end{split}
\end{equation}\\
\indent The Bartlett-corrected likelihood ratio $w_\mu^\ast = w_\mu / \text{E}[w_\mu]$ can be estimated numerically using the Lawley formula defined by Eq.~\eqref{eq:lawelyComposite_correction_def}, which predicts the expectation value of $w_\mu$ to be
\begin{equation}
\label{eq:lawley_GVMaverage_mu}
    \text{E}[w_\mu]= 1+\sum_i^N\mathcal{O}(\varepsilon_i^4)\,.
\end{equation}\\
Therefore, the Bartlett correction factor is equal to unity up to $\sum_i\mathcal{O}(\varepsilon_i^4)$, indicating that any deviations of the likelihood ratio's density function from its asymptotic distribution are also expected to be $\sum_i\mathcal{O}(\varepsilon_i^4)$. 

To further improve the accuracy of the Lawley formula, one can compute the Bartlett correction numerically. Specifically, one can approximate the expectation value of $w_\mu$ by generating data with all the model parameters set to their maximum likelihood estimates
and treating it as a constant independent of the model parameters:
\begin{equation}
    \text{E}[w_\mu]\simeq\text{E}[w_{\hat{\mu}}]\,.
\end{equation}
This approximation has been found to yield highly accurate results, as shown below, and moreover it significantly speeds up the computation of confidence intervals. Rather than generating a new set of data to estimate the Bartlett correction for every tested value of $\mu$, data only needs to be generated once.

In certain scenarios, an analyst may wish to conduct inference on one or more nuisance parameters $\theta_i$ for example to generate a ranking plot of the systematics or obtain the correlation matrix of their estimators. In such cases, the nuisance parameters must be treated as parameters of interest. According to the Lawley formula, the expected value of the likelihood ratio is
\begin{equation}
\label{eq:lawley_GVMaverage}
\begin{split}
    \text{E}[w_{\mu,\boldsymbol{\theta}}] =& 1+M-4\sum_i^M k^{\theta_i\theta_i}\frac{\varepsilon_i^2}{v_i}\\
    &-\sum_i^M\left(k^{\theta_i\theta_i}\right)^2\frac{\varepsilon_i^2}{v_i^2}+\sum_i^N\mathcal{O}(\varepsilon_i^4)\,,
\end{split}
\end{equation}
where, $M$ represents the number of nuisance parameters that have been promoted to parameters of interest. The term $k^{\theta_i\theta_i}$ refers to the $\theta_i$ component of the inverse of the expectation values of the Hessian matrix of the likelihood, as defined by the first term of equation Eq.~\eqref{eq:k-terms1_def}), and is evaluated at $\sigma^2_{u_i}=v_i$. 

\subsection{Confidence regions}
\label{sec:fullgvmcr}

\begin{figure*}[t]
    \begin{center}
    \includegraphics[width=0.6\textwidth]{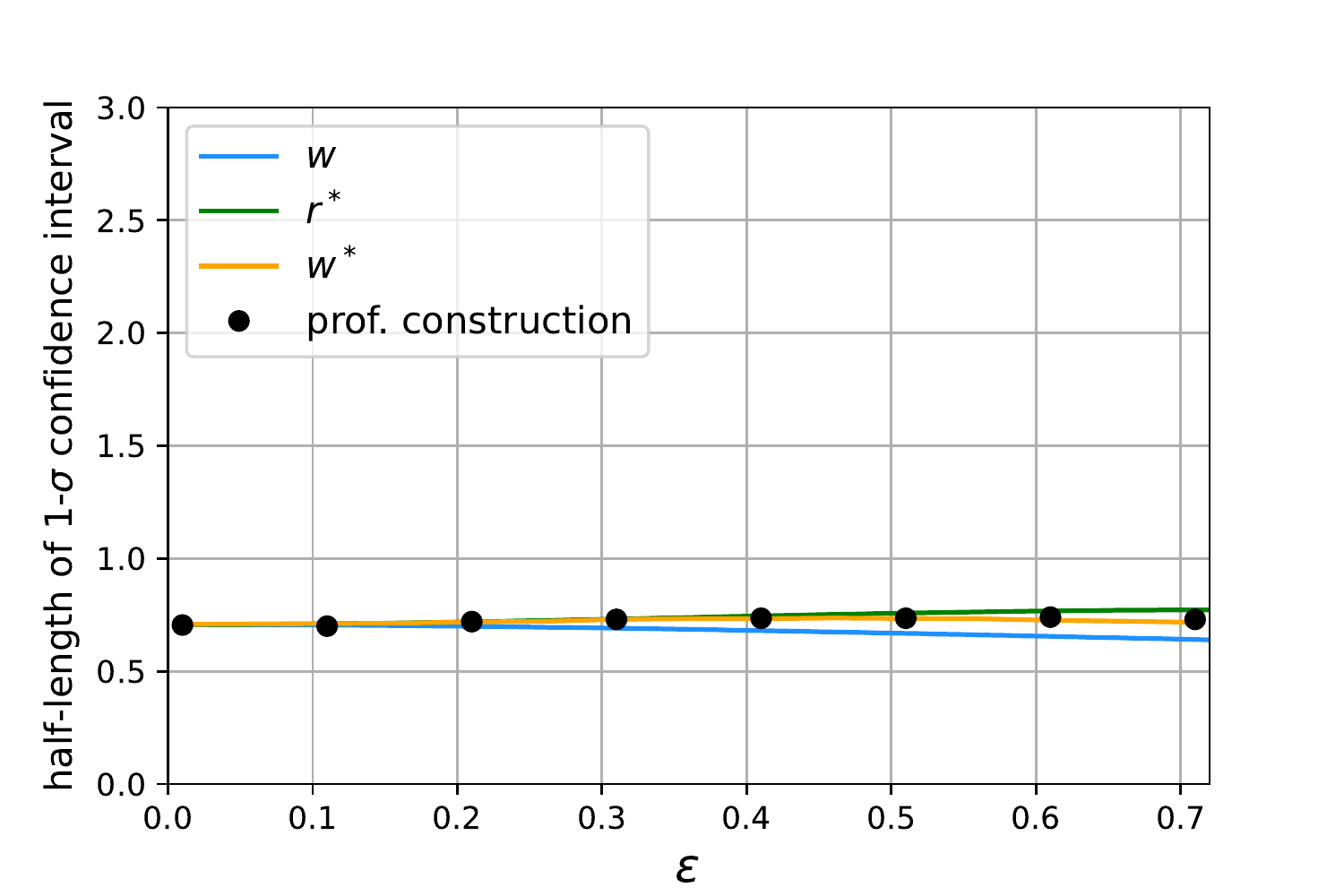}
    \includegraphics[width=0.6\textwidth]{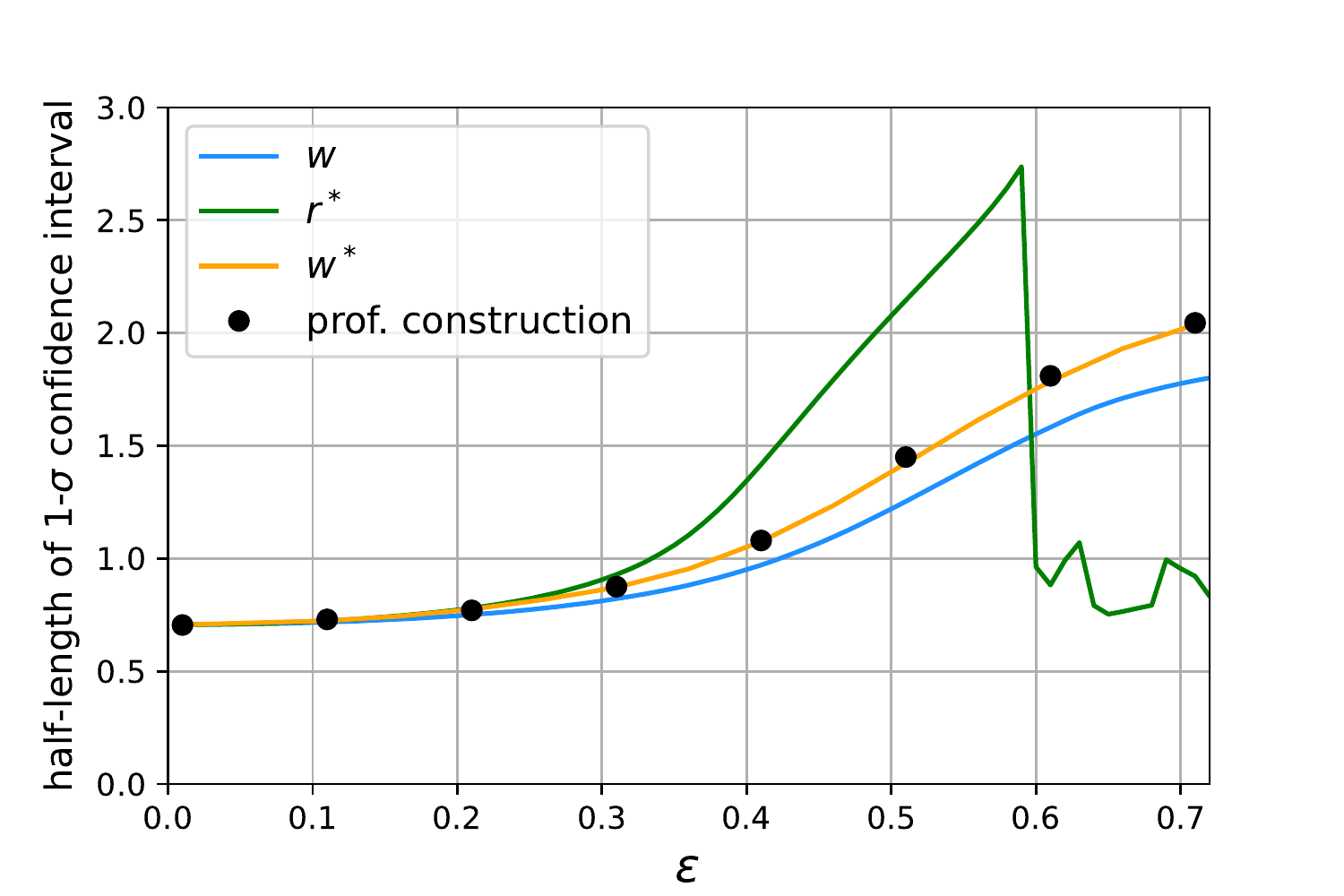}
    \caption{Half-length of $1$-$\sigma$ confidence intervals for parameter $\mu$ (Eq.~\eqref{eq:GVMaverage_logLik}) as a function of $\varepsilon$ for $\delta=0.5$ (top) and $\delta=1.5$ (bottom), computed using $w_\mu$ (blue), $r^\ast_\mu$ (green) and $w^\ast_\mu$ calculated with MC (orange). The black dots represent our most precise estimate of the interval, computed using the profile construction. The confidence interval obtained with $r^\ast_\mu$ breaks down for large values of $\varepsilon$ when there is significant tension in the dataset (bottom).}
    \label{fig:GVMaverage_CI}
    \end{center}
\end{figure*}
As in the previous examples, we compute confidence intervals for the parameter of interest $\mu$ using the likelihood ratio and higher-order statistics, assuming their density functions are given by the asymptotic distributions. Specifically, consider an example similar to what was used in Sec.~\ref{sec:sam}, namely, the mean of two measurements, $y_1=-\delta$ and $y_2=+\delta$, here with associated statistical errors $\sigma_1$ and $\sigma_2$ both equal to $1/\sqrt{2}$ and $\delta=0.5$ or $1.5$. Additionally, we assume that the control measurements $u_1$ and $u_2$ have observed values of $0$, and the estimates of the systematic errors to be $1/\sqrt{2}$, or equivalently, the estimates of the variances $v_1$ and $v_2$ to be $1/2$. Both measurements are assumed to have equal error on error parameters, $\varepsilon_1=\varepsilon_2=\varepsilon$, and we look at the results for different $\varepsilon$.

Figure~\ref{fig:GVMaverage_CI} shows the confidence interval  for the parameter $\mu$ found using the likelihood ratio $w_\mu$, as well as from the higher-order statistics $w_\mu^\ast$ and $r_\mu^\ast$. The resulting confidence intervals are compared to what is found using the profile construction method, which is taken as the best available estimate of such intervals. Among the three methods, $w_\mu^\ast$ is  the most accurate, almost perfectly overlapping with the numerical predictions obtained using the profile construction method. Moreover, $w_\mu^\ast$ is significantly faster to compute as it only requires data generation for $\mu=\hat{\mu}$. In contrast, the profile construction method entails generating a new set of data for every tested value of $\mu$. 

\begin{figure*}[t]
    \begin{center}
    \includegraphics[width=0.6\textwidth]{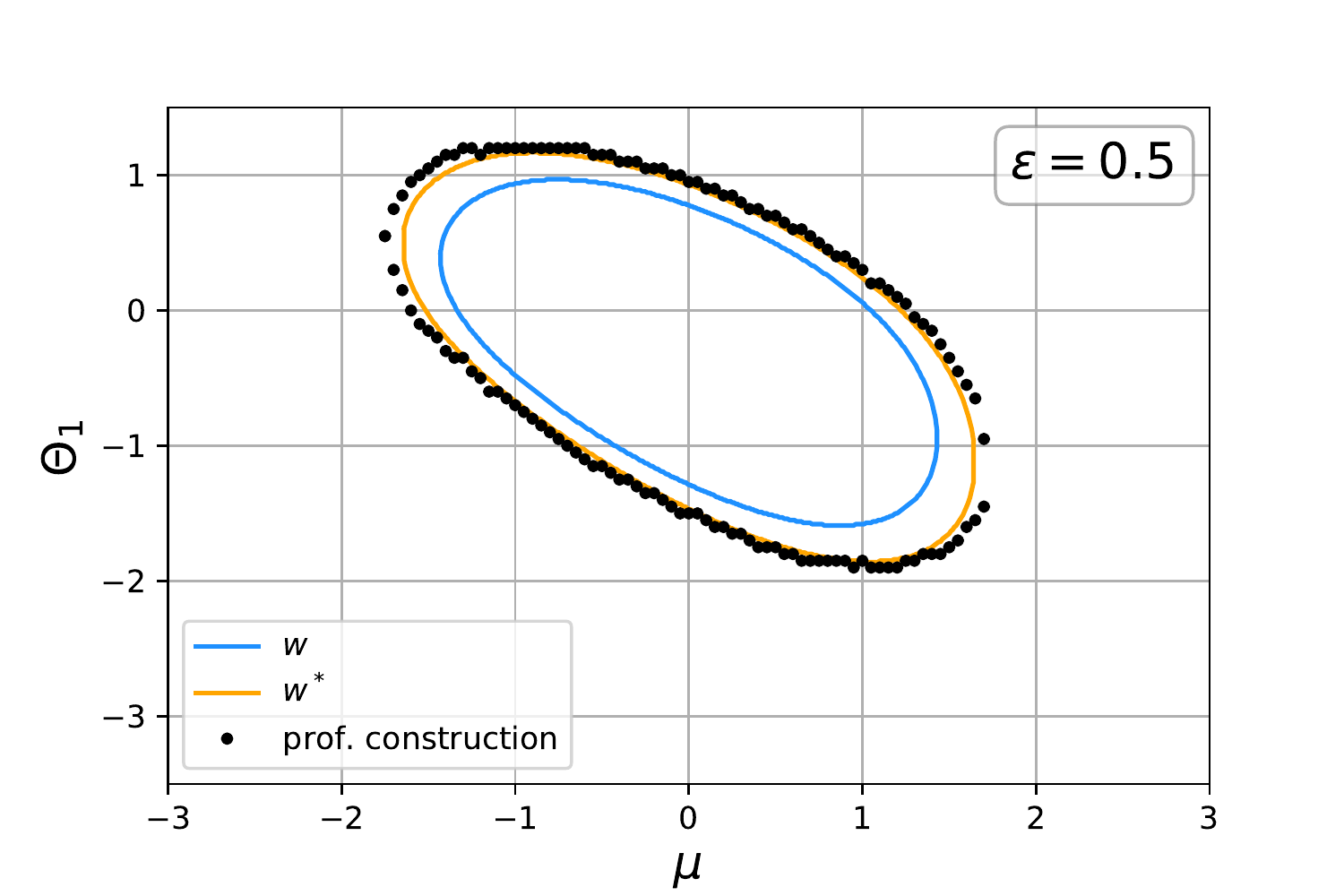}
    \includegraphics[width=0.6\textwidth]{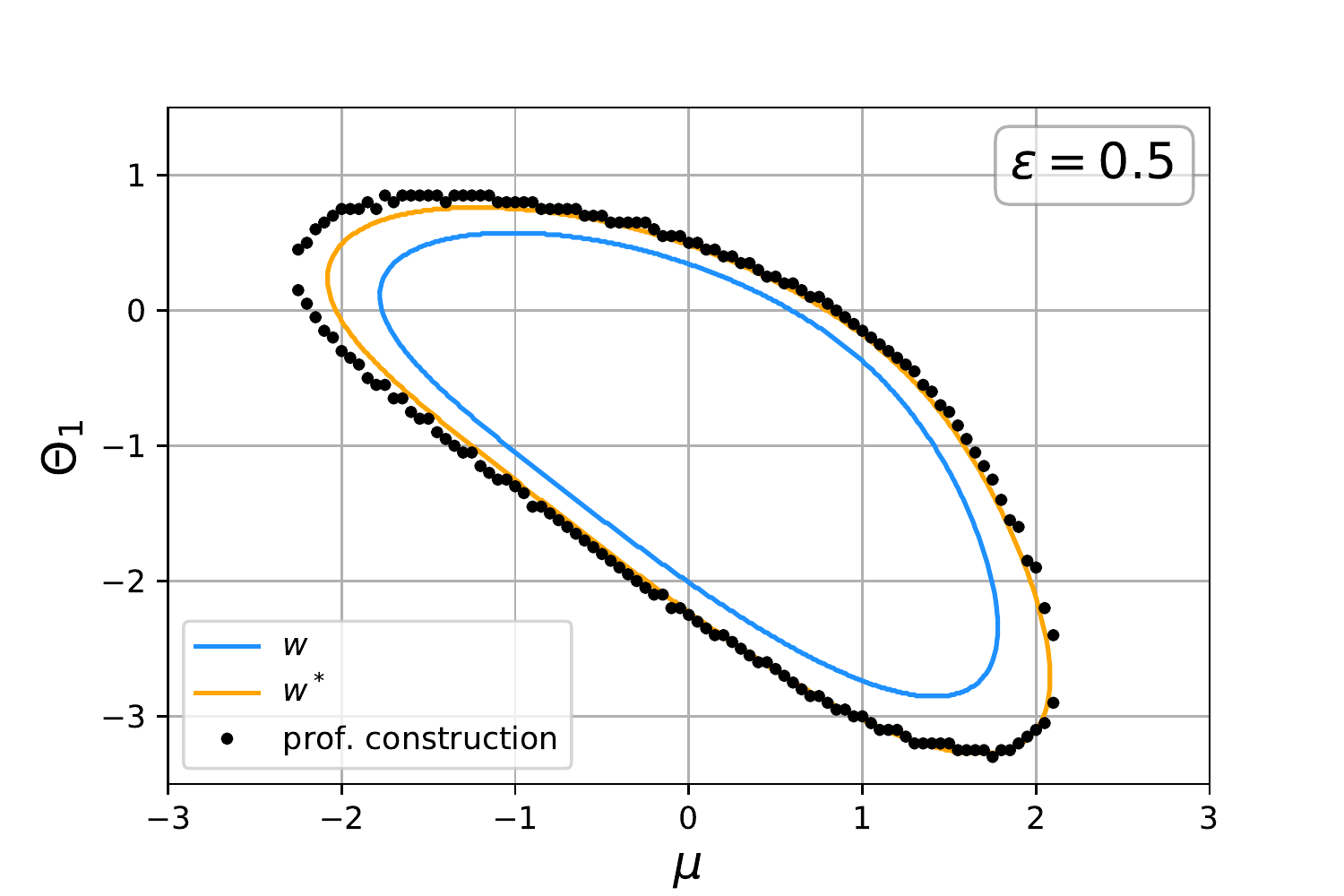}
    \caption{$68.3\%$ confidence regions in $(\mu, \theta_1)$ plane for $\delta=0.5$ (top) and $\delta=1.5$ (bottom), computed using the likelihood ratio (blue) and the Bartlett correction calculated with the Lawley formula (orange). The black dots represent the most precise estimate of the interval, computed using the profile construction. The error-on-error parameters $\varepsilon_1$ and $\varepsilon_2$ are fixed to $0.5$.}
    \label{fig:GVMaverage_CI2}
    \end{center}
\end{figure*}

The $r_\mu^\ast$ statistic, on the other hand, provides accurate predictions for internally consistent data (see the top plot of Fig.~\ref{fig:GVMaverage_CI}). However, for larger discrepancies between the measurements (bottom plot of Fig.~\ref{fig:GVMaverage_CI}), it is reliable only for small values of $\varepsilon$. In this example, it starts deviating from the numerical prediction for $\varepsilon$ above $0.3$ and breaks down for $\varepsilon$ exceeding $0.6$, resulting in numerical instabilities. To assess the usability of $r_\mu^\ast$, one can check whether the logarithmic terms in the log-likelihood associated with the control measurements $\boldsymbol{u}$ fulfill the perturbative condition given by Eq.~\eqref{eq:conv_radius} for the endpoints of the confidence interval. For $\delta=1.5$, this condition limits the applicability of $r_\mu^\ast$ to $\varepsilon\simeq0.3$, whereas, for $\delta=0.5$, the threshold is higher, above $0.6$.\\
\indent Figure \ref{fig:GVMaverage_CI2} illustrates the $2$D confidence regions in the $(\mu,\theta_1)$ plane. These confidence regions were computed by fixing $\theta_2$ to its profiled value, while treating $\theta_1$ as a parameter of interest. A similar exercise could have been conducted for $\mu$ and $\theta_2$, or $\theta_1$ and $\theta_2$. The results shown in Fig.~\ref{fig:GVMaverage_CI2} were derived using the same measured data as in the previous example, with the error on error parameters $\varepsilon_1$ and $\varepsilon_2$ fixed at $0.5$. The confidence regions were computed using the likelihood ratio $w_{\mu,\boldsymbol{\theta}}$ and the Bartlett-corrected likelihood ratio $w_{\mu,\boldsymbol{\theta}}^\ast$ computed via Eq.~\eqref{eq:lawley_GVMaverage}, and were then compared with the confidence regions estimated using the profile construction technique. The results indicate that the Bartlett correction improves the accuracy of predictions significantly compared to the uncorrected likelihood ratio.
\subsection{Goodness of fit}
\label{sec:GVM_GoF}
The goodness-of-fit statistic for the Gamma Variance Model can be defined using the same approach as used for the simple-average model in Sec.~\ref{sec:simpleAverage_gof}, leading to 
\begin{equation}
\begin{aligned}
    q=-2\log L(\hat{\mu},\hat{\boldsymbol{\theta}},\widehat{\widehat{\boldsymbol{\sigma_u^2}}})
    = \sum_{i=1}^N \left[ \frac{(y_i-\hat{\mu}-\hat{\theta}_i)^2}{\sigma_{y_{i}}^2}\right.\\
    \left.+\left(1+\frac{1}{2\varepsilon_i^2}\right)\log\left( 1+2\varepsilon_i^2\frac{(u_i-\hat{\theta}_i)^2}{v_i}
    \right) \right]\,.
\end{aligned}
\end{equation}
In this case, however, constructing a saturated model is not useful because the Lawley formula gives $b=0$ at order $\varepsilon_i^2$. Nonetheless, the Bartlett correction can still be computed using MC. This method allows for a significant computational improvement over generating the exact distribution of $q$ using pseudo-experiments to estimate its $p$-value. The latter approach would require roughly on the order of $10^5$ simulated experiments to accurately capture a $4\sigma$ effect, while the expectation value of $q$ can be estimated with a precision of several percent using only $\mathcal{O}(10^3)$ pseudo-experiments.
\begin{figure*}[t]
    \begin{center}
    \includegraphics[width=0.6\textwidth]{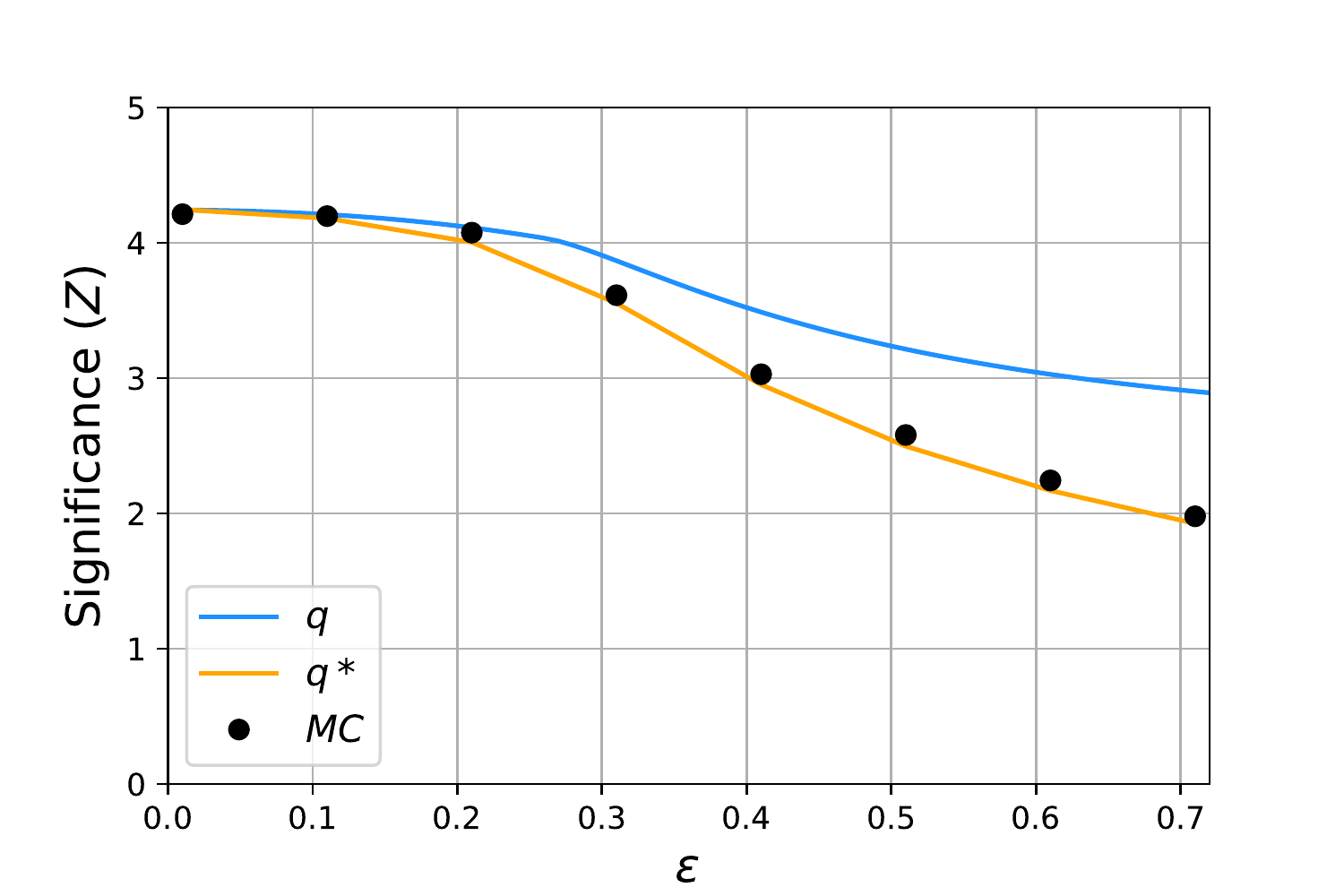}
    \caption{Significance of discrepancy as defined by Eq.~\eqref{eq:significance} plotted as a function of $\varepsilon$. Higher significance values indicate greater tension between the data. The significance was computed using the goodness-of-fit statistic (blue) defined by Eq.~\eqref{eq:simpleAverage_GoF} and its Bartlett corrected counterpart calculated numerically (orange). The black dots represent the significance computed by generating the distribution of $q$ with MC. The plot illustrates that the Bartlett correction enables an accurate approximation of the numerical predictions.}
    \label{fig:GVMaverage_gof}
    \end{center}
\end{figure*}

To illustrate these techniques we compute the significance of the $p$-value for an average of two incompatible measurements using Eq.~\eqref{eq:GVMaverage_logLik}. The observed values of $y_1$ and $y_2$ are assumed to be $-3$ and $3$ whereas the control measurements $u_1$ and $u_2$ are set to $0$. The statistical uncertainties, $\sigma_1$ and $\sigma_2$, are set to $1/\sqrt{2}$ as the estimates of the systematic errors (which is equivalent to set $v_1=v_2=1/2$). Figure~\ref{fig:GVMaverage_gof} compares the significance computed using the goodness-of-fit statistic $q$ and the Bartlett-corrected $q^\ast$, with the significance computed generating the distribution of $q^\ast$ numerically, all of them as a function of $\varepsilon$. Consistent with our earlier findings, the Bartlett correction is found to yield highly accurate predictions.

\section{Conclusions}
\label{sec:conclusions}

We have demonstrated the efficacy of higher-order asymptotics in the computation of confidence intervals and $p$-values within the framework of the Gamma Variance Model, a specialized statistical model designed to address uncertainties in parameters that themselves represent uncertainties. The methods studied in this paper hold particular relevance when the GVM's fixed parameters $\varepsilon$, indicative of the relative uncertainties in estimates of standard deviations for Gaussian-distributed measurements, are not negligible. In such scenarios, standard asymptotic methods are unable to provide accurate confidence intervals or $p$-values.

Our investigation specifically focused on the $r^\ast$ statistic and the Bartlett correction, both of which are higher-order asymptotic techniques that offer adjustments to the first-order (profile) likelihood ratio and likelihood root test statistics. These adjustments enable the test statistics to be more accurately approximated by their asymptotic distributions, even when the $\varepsilon$ parameter is large.

Both the Barndorff-Nielsen $r^\ast$ statistic and the Bartlett corrected likelihood ratio demonstrated their value as tools to enhance the accuracy and reliability of confidence interval and $p$-value calculations using Gamma Variance Models. However, it should be noted that the $r^\ast$ statistic exhibited instabilities in the presence of internally incompatible data for large values of $\varepsilon$. Additionally, while $r^\ast$ can be computed analytically for all the examples examined in this paper, the expressions become complicated for models associated with realistic applications, such as the simple-average and full GVM models. 

Conversely, the Bartlett correction, calculated using the Lawley formula \eqref{eq:lawelyComposite_correction_def}, offers a more elegant expression for the expectation value of the likelihood ratio, which is employed to compute the Bartlett correction factor for the likelihood ratio. Specifically, refer to Eq.\eqref{eq:lawley_sam} for the simple-average model (see Sec.\ref{sec:sam}) and Eqs.\eqref{eq:lawley_GVMaverage_mu} and~\eqref{eq:lawley_GVMaverage} for combinations utilizing the full GVM (see Sec.~\ref{sec:fullgvmcr}). In addition, when necessary, estimating the Bartlett correction numerically is straightforward, as shown in Sec.~\ref{sec:GVM_ave}.

The Bartlett correction also proved to be an effective technique for improving the goodness-of-fit statistic. This was true for cases where the statistic could be computed analytically using the Lawley formula, such as the simple-average model (see Eq.~\eqref{eq:lawley_q} in Sec.~\ref{sec:simpleAverage_gof}), as well as for cases where it was estimated using Monte Carlo methods, as in the full Gamma Variance Model (see Sec.~\ref{sec:GVM_GoF}). The application of the Bartlett correction in the latter scenario significantly reduced the number of pseudo-experiments required for accurately estimating the significance of rare effects.

Overall, to improve both the applicability and precision of the GVM, the Bartlett correction has proven to be a more reliable and versatile solution. 

Furthermore, these findings highlight the potential of higher-order asymptotics to refine inference on the parameters of interest in various contexts, not only the GVM. Higher-order asymptotics are valuable tools when the MLEs of statistical models do not follow Gaussian distributions or, equivalently, when log-likelihoods are not well approximated by quadratic expressions. For Gamma Variance Models this occurs when $\varepsilon$ is large; however, in general, such deviations are typically associated with small experimental sample sizes.  In Particle Physics, it is not uncommon to search for new
signal processes by counting collision events with very specific characteristics, such that the expected number of background events may be order unity.  With sample sizes of this order it is expected that asymptotic distributions will not be accurate and high-order asymptotic formulae should prove valuable (see, e.g., Refs.~\cite{bib:Brazzale2007,bib:Davison2008}.  

The introduction of higher-order asymptotic corrections removes a potential stumbling block for use of the Gamma Variance Model.  As many estimates of systematic uncertainties may themselves be uncertain at the level of 20\% to 50\% or more, one would not expect asymptotic confidence intervals or $p$-values to be accurate.  By using higher-order corrections, accurate results can be achieved without minimal or no Monte Carlo simulation, greatly simplifying use of the model.

\section*{Acknowledgements}

EC and GC are grateful to Bogdan Malaescu and other colleagues in the ATLAS Collaboration
who have provided valuable feedback on this work and to the
U.K.\ Science and Technology Facilities Council for its support. ARB was supported by the Project of Excellence (2018-2022)
of the Department of Statistical Sciences of the University
of Padova. The authors are grateful to the organisers of the BIRS Workshop ``Systematic Effects and Nuisance Parameters in Particle Physics Data Analyses'' in Banff, where productive discussions on this work took place.


\begin{thebibliography}{[88]}
\bibitem{bib:Wilks1938} S.S.~Wilks, {\it The large-sample distribution 
of the likelihood ratio for testing composite hypotheses}, Ann.\ Math.\
Statist.\ {\bf 9} (1938) 60-2.
\bibitem{bib:Wald1943} A.~Wald, {\it Tests of Statistical Hypotheses
Concerning Several Parameters When the Number of Observations is
Large}, Transactions of the American Mathematical Society, Vol.~{\bf
54}, No.~3 (Nov., 1943), pp.~426-482.
\bibitem{bib:Cowan2011} G.~Cowan, K.~Cranmer, E.~Gross and O.~Vitells, 
{\it Asymptotic formulae for likelihood-based tests of new physics}, Eur.\ Phys.\ J.\ C 71 (2011) 1554.
\bibitem{bib:Algeri2019}
S.~Algeri, J.~Aalbers, K.~Dundas Mor\r{a} and J.~Conrad,
{\it Searching for new phenomena with profile likelihood ratio tests},
Nature Rev. Phys. \textbf{2} (2020) no.5, 245-252
doi:10.1038/s42254-020-0169-5
\bibitem{bib:Barndorff-Nielsen1980} O. Barndorff-Nielsen,
{\it Conditionality Resolutions}, Biometrika , 67 (1980) 2, pp. 293-310.
\bibitem{bib:Bartlett1937}  M.S.~Bartlett, {\it 
Properties of sufficiency and statistical tests}, 
Royal Society of London Proceedings Series A 160, (1937) 268-282.
\bibitem{bib:Brazzale2007} A.R.~Brazzale, A.C.~Davison and N.~Reid, {\it Applied Asymptotics: Case Studies in Small-Sample Statistics},
  Cambridge University Press (2007).
\bibitem{bib:Cordeiro2014} Gauss M.~Cordeiro and Francisco Cribari-Neto,
{\it An Introduction to Bartlett Correction and Bias Reduction},
Springer Verlag, 2014.
\bibitem{bib:Cowan2019} G. Cowan, {\it Statistical Models with Uncertain Error Parameters}, Eur. Phys. J. C (2019) 79:133; arXiv:1809.05778.
\bibitem{bib:Xia2021} Li-Gang Xia, {\it Improved Asymptotic Formulae for Statistical Interpretation Based on Likelihood Ratio Tests}, arXiv:2101.06944 [physics.data-an] (2021).
\bibitem{bib:Barndorff-Nielsen1983} O.~Barndorff-Nielsen, {\it On a formula for the distribution of the maximum likelihood estimator}, Biometrika, 70 (2) 1983, pp.~343–365, doi.org/10.1093/biomet/70.2.343.
\bibitem{bib:Barndorff-Nielsen1986} O.~Barndorff-Nielsen, {\it Inference on full or partial parameters based on the standardized signed log likelihood ratio}, Biometrika, Volume 73, Issue 2, August 1986, pp.~307–322.
\bibitem{bib:Barndorff-Nielsen1990} Barndorff-Nielsen, O.E. (1990), {\it Approximate Interval Probabilities.}, Journal of the Royal Statistical Society: Series B (Methodological), 52: 485-496. 
\bibitem{bib:Davison2022} Anthony C. Davison and Nancy Reid, {\it The Tangent Exponential Model}, 2022, arXiv:2106.10496v2
\bibitem{bib:Davison2006} A. C. Davison and others, {\it Improved Likelihood Inference for Discrete Data}, Journal of the Royal Statistical Society Series B: Statistical Methodology, Volume 68, Issue 3, June 2006, Pages 495–508
\bibitem{bib:Lawley1956}  D.N.~Lawley, {\it A general method for
approximating to the distribution of likelihood ratio criteria}, 
Biometrika, Vol.\ 43, Issue 3-4, (1956) 295-303.
\bibitem{bib:Cowan2022} G.~Cowan, {\it Effect of Systematic Uncertainty Estimation on the Muon $g-2$ Anomaly}, EPJ Web of Conferences 258, 09002 (2022); arXiv:2107.02652.
\bibitem{bib:Dose2014}   Wolfgang von der Linden, Volker Dose and
Udo von Toussaint, {\it Bayesian Probability Theory:
Applications in the Physical Sciences}, Cambridge University Press, 2014.
\bibitem{bib:Dagostini1999}   G. D'Agostini, {\it Sceptical combination
    of experimental results: General considerations and application to
    epsilon-prime/epsilon}, arXiv:hep-ex/9910036 (1999).
\bibitem{bib:Cowan2006}  G.~Cowan, {\it Bayesian Statistical Methods for
    Parton Analyses}, in {\it Proceedings of the 14th International
    Workshop on Deep Inelastic Scattering (DIS2006)}, M.~Fuzz,
  K.~Nagano, and K.~Tokushuku (eds.), Tsukuba, 2006.
\bibitem{bib:Erler2020}  Jens Erler and Rodolfo Ferro-Hern\'andez, {\it Alternative to the application of PDG scale factors},  Eur.\ Phys.\ J.\ C 80 (2020) 6, 541, arXiv:2004.01219.
\bibitem{bib:Cranmer2005} K.~Cranmer, {\it Statistical challenges for searches for new
physics at the LHC}, in Proceedings of PHYSTAT05, L.~Lyons and M.K.~Unel (eds.), Imperial College Press, pp. 112-123 (2005)
\bibitem{bib:Chuang2000} C.~Chuang and T.L.~Lai, {\it Hybrid resampling methods for
confidence intervals}, Statistica Sinica 10 (2000) 1-50.
\bibitem{bib:BodhisattvaSen2009} Bodhisattva Sen, Matthew Walker and Michael Woodroofe, {\it On the Unified Method with Nuisance Parameters}, Statistica Sinica 19 (2009) 301-314.
\bibitem{bib:Davison2008}  A.C.~Davison and N.~Sartori, {\it The Banff Challenge: Statistical Detection of a Noisy Signal}, Statistical Science, Vol. 23, No. 3 (2008), pp. 354-364.

\end{thebibliography}
\end{document}